%% file: main.tex
\documentclass[acmtog,nonacm]{acmart}
\acmSubmissionID{539}

\usepackage{booktabs} 

\usepackage[ruled]{algorithm2e} 

\SetAlFnt{\small}
\SetAlCapFnt{\small}
\SetAlCapNameFnt{\small}
\SetAlCapHSkip{0pt}

\usepackage{subfig}
\usepackage{multirow}
\newcommand{\tabincell}[2]{\begin{tabular}{@{}#1@{}}#2\end{tabular}} 
\usepackage{comment}
\usepackage{xspace}

\makeatletter
\DeclareRobustCommand\onedot{\futurelet\@let@token\@onedot}
\def\@onedot{\ifx\@let@token.\else.\null\fi\xspace}

\def\eg{\emph{e.g}\onedot}

\def\etal{\emph{et al}\onedot}
\makeatother

\usepackage{amsmath}
\DeclareMathOperator*{\argmax}{arg\,max}

\acmJournal{TOG}

\setcopyright{acmcopyright}\acmJournal{TOG}
\acmYear{2022}\acmVolume{41}\acmNumber{4}\acmArticle{83}\acmMonth{7} \acmDOI{10.1145/3528223.3530148}
\begin{document}
\title{Semantically Supervised Appearance Decomposition for Virtual Staging from a Single Panorama
}

\author{Tiancheng Zhi}
\orcid{0000-0002-0953-1444}
\affiliation{%
\institution{Carnegie Mellon University}
\country{USA}}
\email{zhitiancheng@gmail.com}
\author{Bowei Chen}
\orcid{0000-0002-2225-8796}
\affiliation{%
\institution{Carnegie Mellon University}
\country{USA}}
\email{boweiche@andrew.cmu.edu}
\author{Ivaylo Boyadzhiev}
\orcid{0000-0001-8555-2952}
\affiliation{%
\institution{Zillow Group}
\country{USA}}
\email{ivaylob@zillowgroup.com}
\author{Sing Bing Kang}
\orcid{0000-0003-2016-2915}
\affiliation{%
\institution{Zillow Group}
\country{USA}}
\email{singbingk@zillowgroup.com}
\author{Martial Hebert}
\orcid{0000-0003-4566-5930}
\affiliation{%
\institution{Carnegie Mellon University}
\country{USA}}
\email{mhebert@andrew.cmu.edu}
\author{Srinivasa G. Narasimhan}
\orcid{0000-0003-0389-1921}
\affiliation{%
\institution{Carnegie Mellon University}
\country{USA}}
\email{srinivas@andrew.cmu.edu}

\renewcommand\shortauthors{Zhi, T. et al}


\input{fig_teaser}
\input{sec0_abstract}

%
%

%
%


\maketitle
\input{fig_framework}

{
\renewcommand{\thefootnote}{\fnsymbol{footnote}}
\footnotetext[1]{3D model credits: cake\copyright HQ3DMOD/Adobe Stock, 
lounge\copyright JaskaranSingh/Adobe Stock, 
cherries\copyright VIZPARK/Adobe Stock, 
table\copyright Claudio Naviglio/Adobe Stock, 
tea\copyright Brandon Westlake/Adobe Stock, dining 
set\copyright Oleg Kuch/Adobe Stock,
dinnerware\copyright HQ3DMOD/Adobe Stock,
utensil\copyright Dmitrii Ispolatov/Adobe Stock,
pot\copyright Brandon Westlake/Adobe Stock, side 
table\copyright JaskaranSingh/Adobe Stock, 
frames\copyright Kamil/Adobe Stock, 
sofa\copyright HQ3DMOD/Adobe Stock,
envelope\copyright adobestock3d/Adobe Stock, 
pillow\copyright TurboSquid/Adobe Stock,
beanbag\copyright FongAudishondra/Adobe Stock,
stand\copyright Dmitrii Ispolatov/Adobe Stock,
plant\copyright Kids Creative Agency/Adobe Stock,
basket\copyright adobestock3d/Adobe Stock.}
}

\input{sec1_intro}

\input{sec2_related}
\input{sec4_semantic}
\input{sec5_detection}
\input{sec6_removal}
\input{sec7_eval}

\input{sec8_apps}

\input{sec9_conclude}

\begin{acks}
This work was supported by a gift from Zillow Group, USA, and NSF Grants \#CNS-2038612,  \#IIS-1900821.
%
%
%
%
\end{acks}

\bibliographystyle{ACM-Reference-Format}
\bibliography{mybib}


\end{document}

%% file: fig_teaser.tex
\begin{teaserfigure}
\centering
\begin{minipage}{0.37\linewidth}
\includegraphics[width=\linewidth]{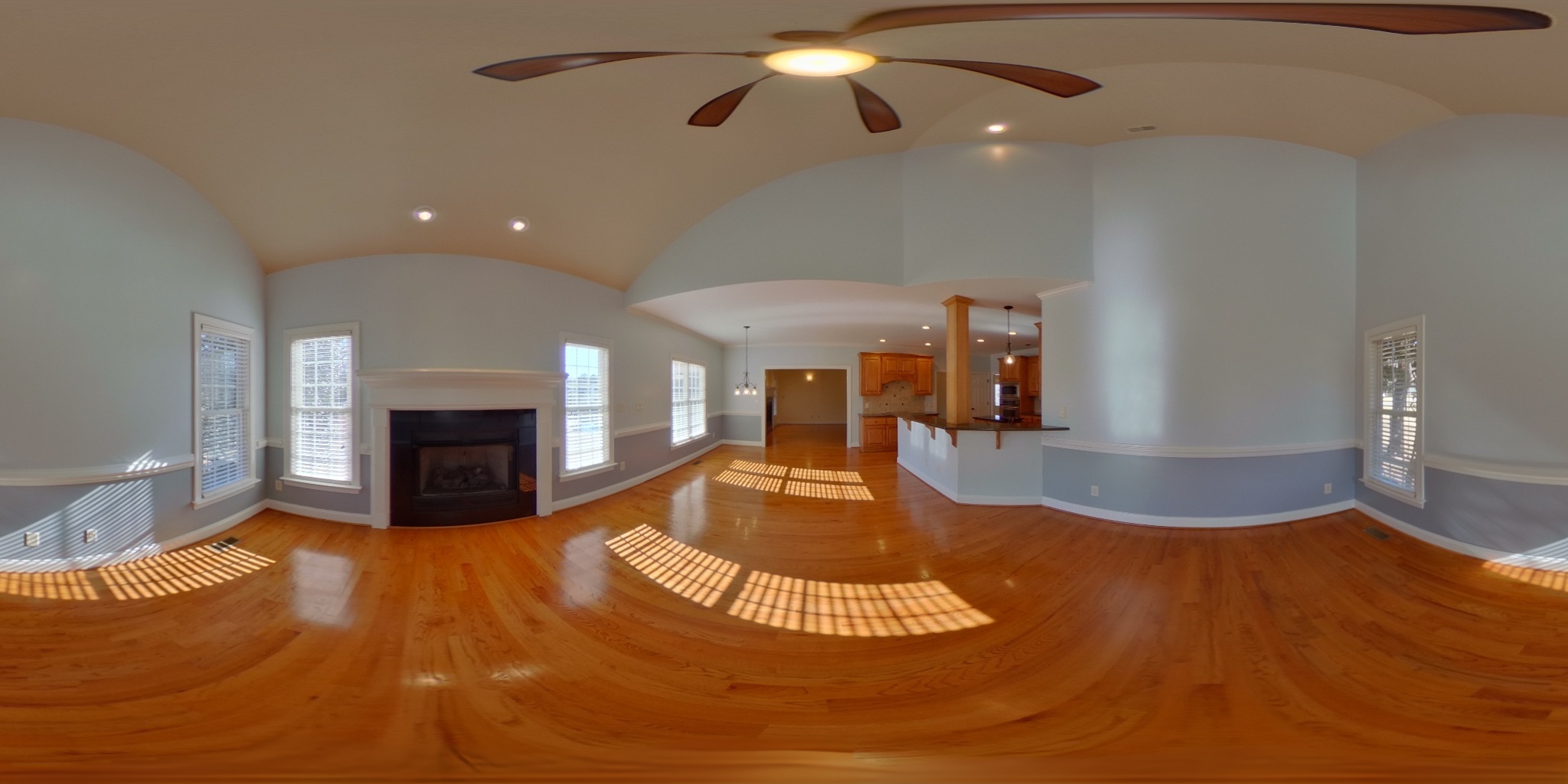}
\end{minipage}
\begin{minipage}{0.122\linewidth}
\includegraphics[width=\linewidth]{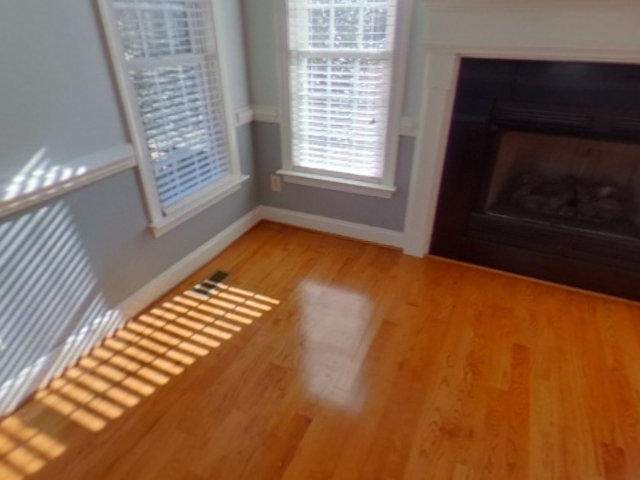} \\
\includegraphics[width=\linewidth]{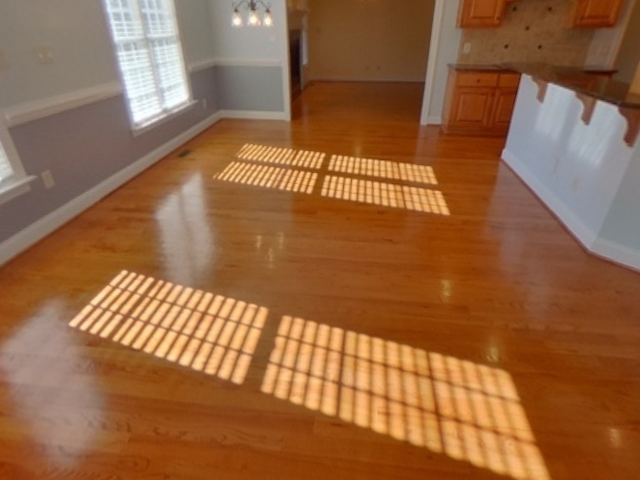}
\end{minipage}
\begin{minipage}{0.37\linewidth}
\includegraphics[width=\linewidth]{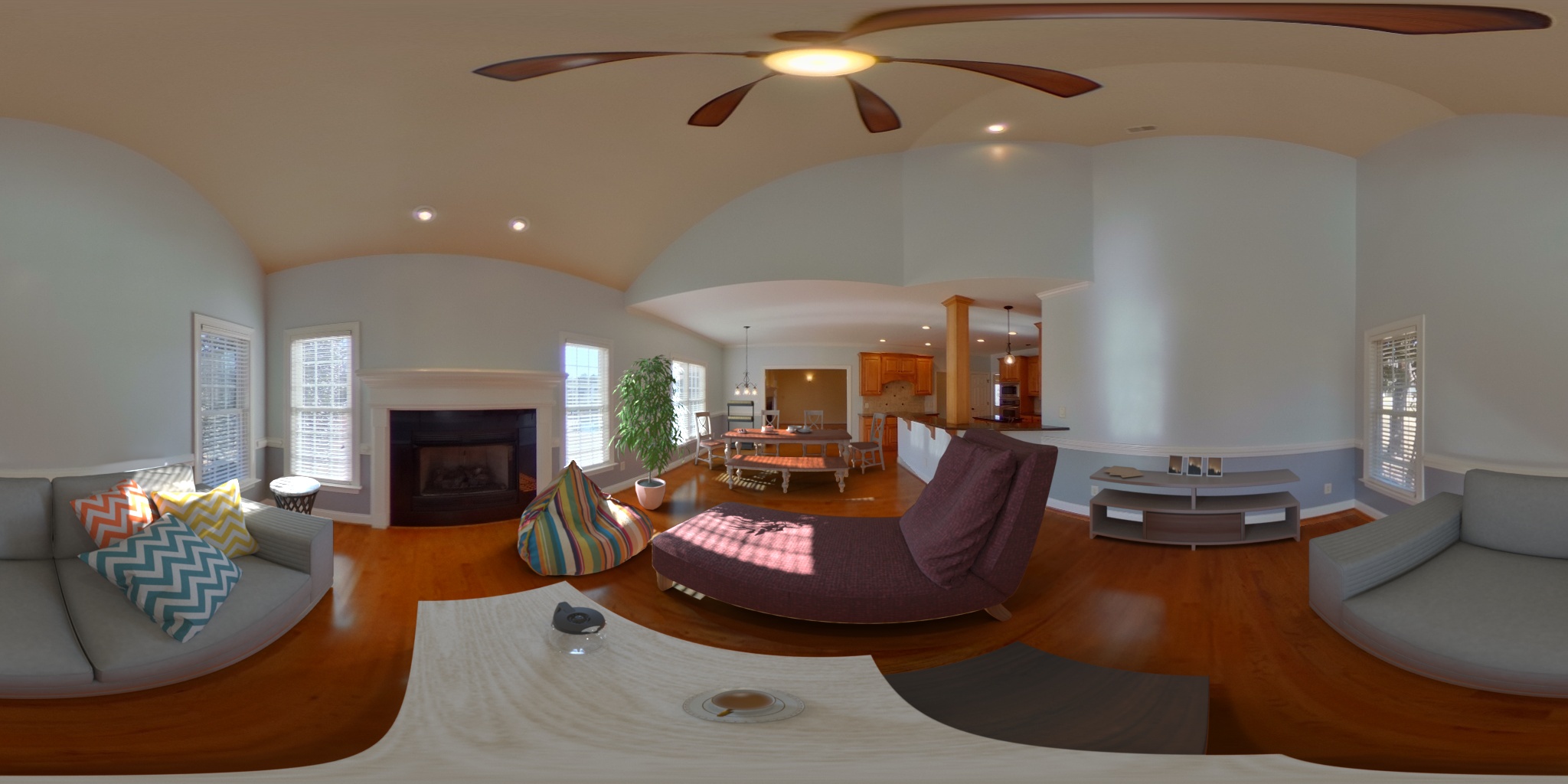}
\end{minipage}
\begin{minipage}{0.122\linewidth}
\includegraphics[width=\linewidth]{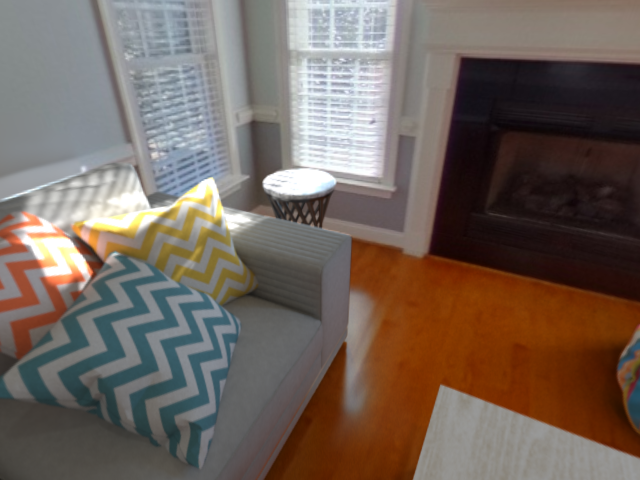} \\
\includegraphics[width=\linewidth]{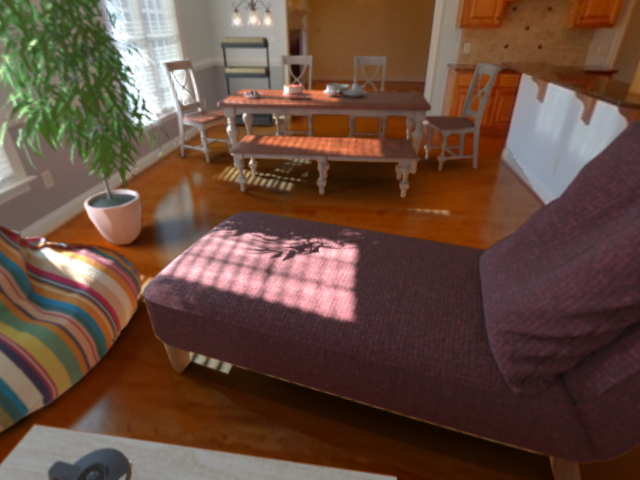}
\end{minipage}
\\
\begin{minipage}{0.37\linewidth}
\includegraphics[width=\linewidth]{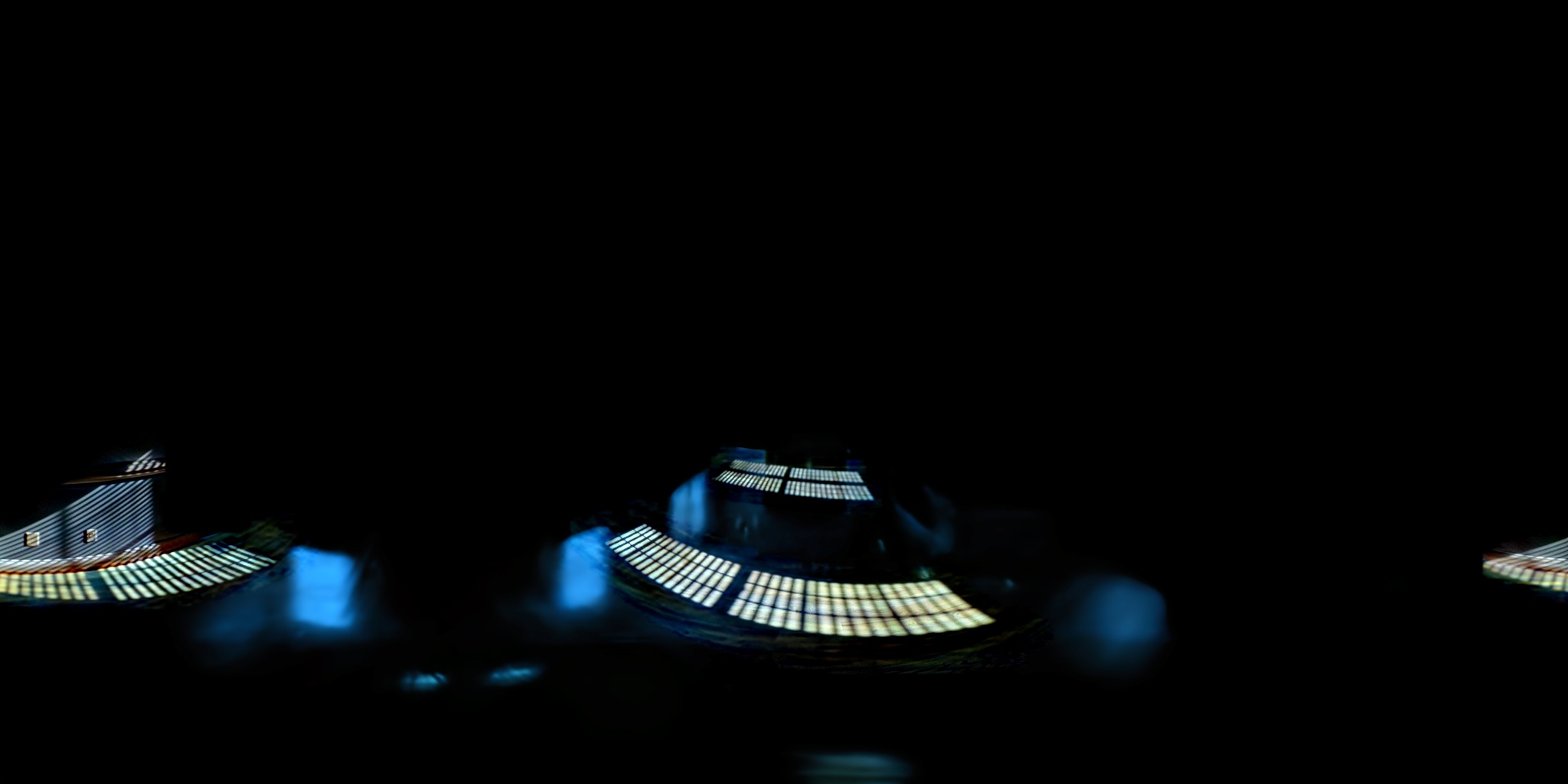}
\end{minipage}
\begin{minipage}{0.122\linewidth}
\includegraphics[width=\linewidth]{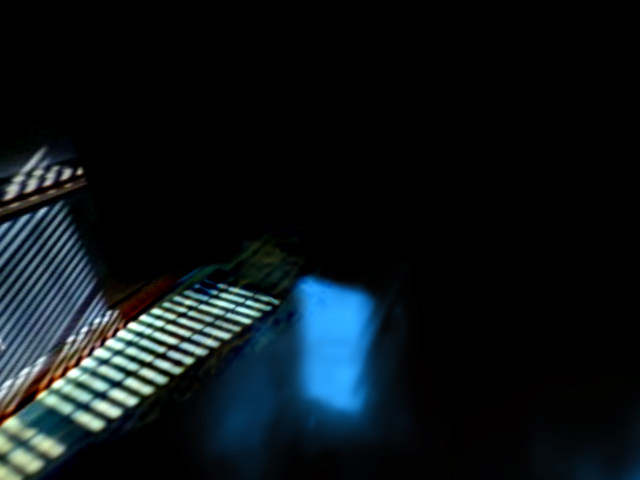} \\
\includegraphics[width=\linewidth]{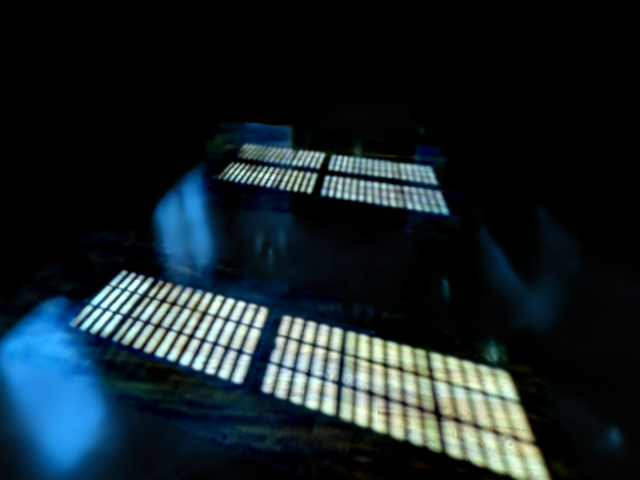}
\end{minipage}
\begin{minipage}{0.37\linewidth}
\includegraphics[width=\linewidth]{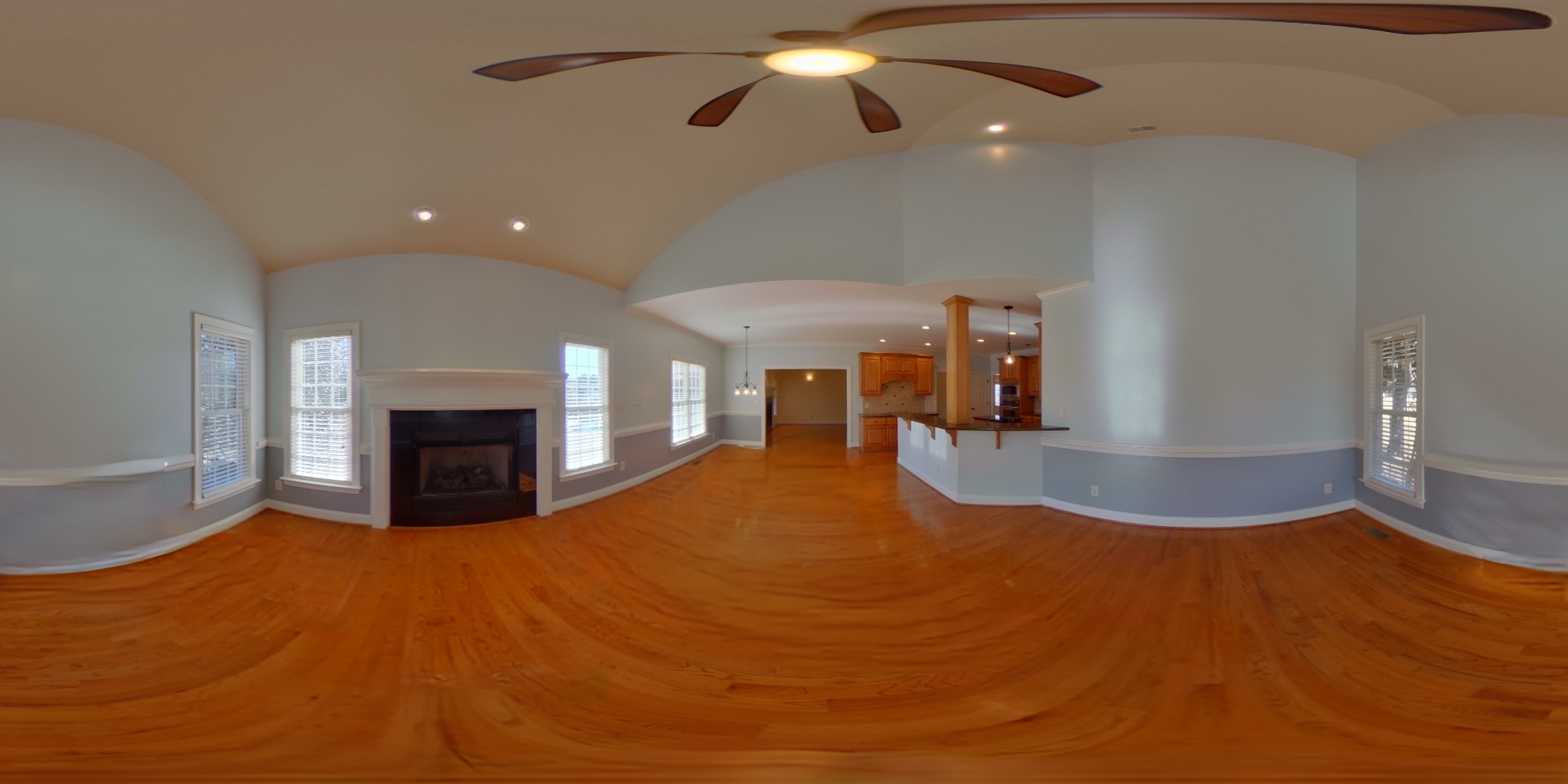}
\end{minipage}
\begin{minipage}{0.122\linewidth}
\includegraphics[width=\linewidth]{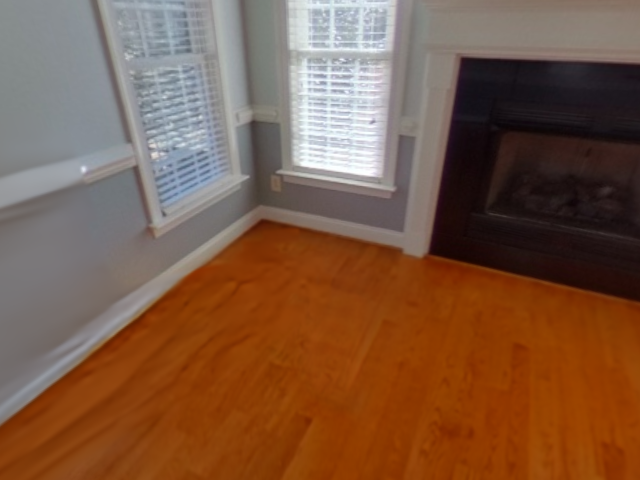} \\
\includegraphics[width=\linewidth]{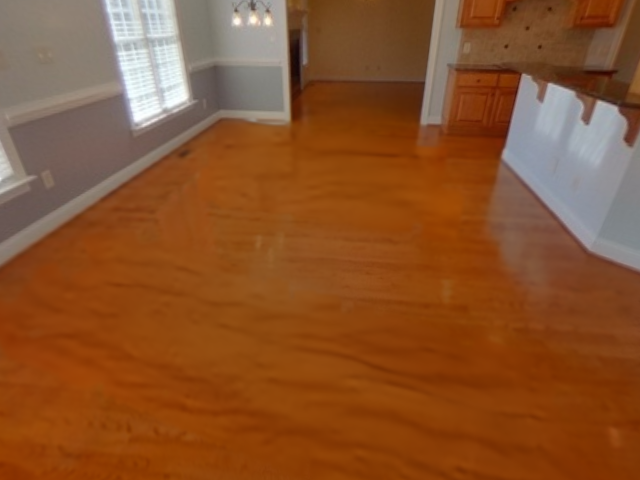}
\end{minipage}
\captionof{figure}{A single panorama of an empty indoor environment (top left) is decomposed into four appearance components on the floor and walls: specular and direct sunlight (bottom left), diffuse (not shown) and diffuse ambient without direct sunlight (bottom right). Although the decomposition is not perfect (e.g. power sockets are removed in the bottom right), we demonstrate its effectiveness in enabling high quality virtual staging applications including furniture insertion\protect\footnotemark (top right), changing flooring and sun direction (Sec.~\ref{sec:apps}). Multiple realistic effects are rendered, including shadows under the table, sunlight on the side table, sofa pillows, beanbag, dining table and chaise lounge, occlusion of specular reflection by the sofa and chair. The plant creates soft and hard shadows by blocking the skylight and the sunlight. On the other hand, inserting furniture without appearance decomposition fails to reproduce these effects (Fig.~\ref{fig:noeff}).}
\label{fig:teaser}
\end{teaserfigure}

%% file: sec0_abstract.tex
\begin{abstract}

We describe a novel approach to decompose a single panorama of an empty indoor environment into four appearance components: specular, direct sunlight, diffuse and diffuse ambient without direct sunlight. Our system is weakly supervised by automatically generated semantic maps (with floor, wall, ceiling, lamp, window and door labels) that have shown success on perspective views and are trained for panoramas using transfer learning without any further annotations. A GAN-based approach supervised by coarse information obtained from the semantic map extracts specular reflection and direct sunlight regions on the floor and walls. These lighting effects are removed via a similar GAN-based approach and a semantic-aware inpainting step. The appearance decomposition enables multiple applications including sun direction estimation, virtual furniture insertion, floor material replacement, and sun direction change, providing an effective tool for virtual home staging. We demonstrate the effectiveness of our approach on a large and recently released dataset of panoramas of empty homes.
\end{abstract}

%% file: fig_framework.tex
\begin{figure*}
    \centering
    \includegraphics[width=0.96\linewidth]{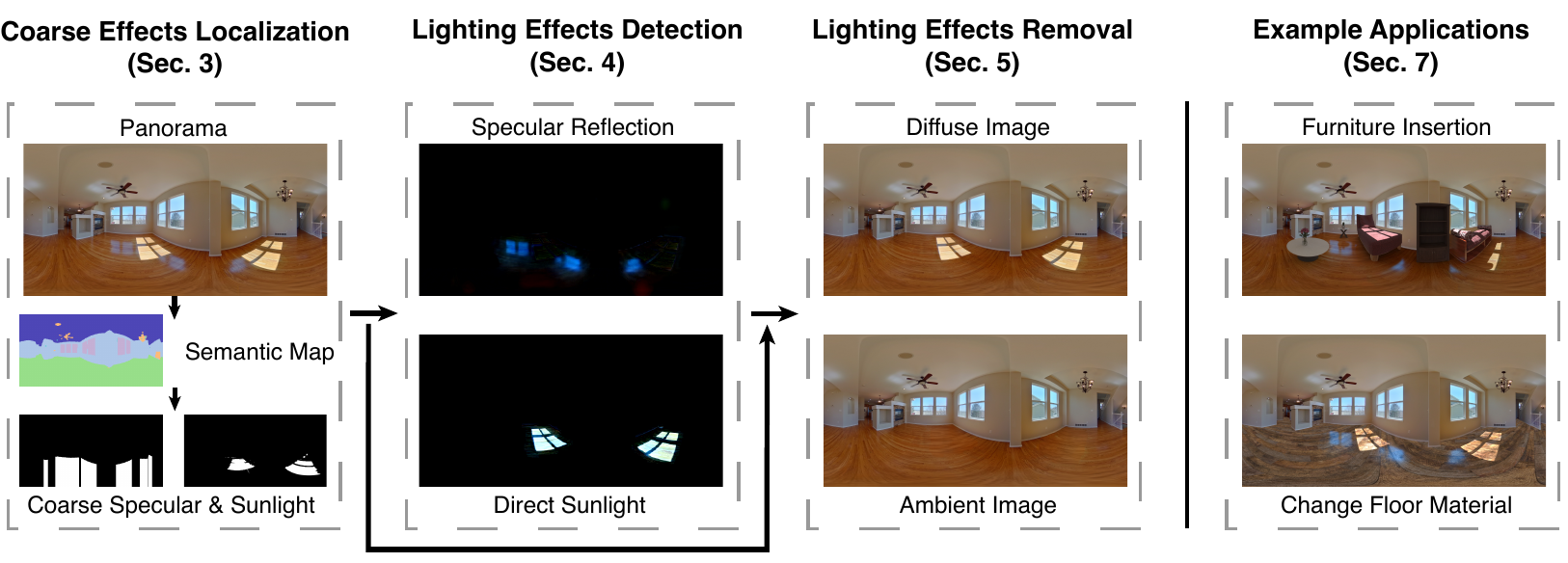}
    \caption{Our system consists of three modules: \emph{Coarse Effects Localization} (Sec.~\ref{sec:semantic}) coarsely localizes lighting effects using automatically generated semantics; \emph{Lighting Effects Detection} (Sec.~\ref{sec:detection}) separates specular reflection on the floor and direct sunlight on the floor and the wall (effects are brightened for visualization); \emph{Lighting Effects Removal}~(Sec. \ref{sec:removal}) removes the detected specular reflection and direct sunlight, outputting a diffuse image (no specular) and an ambient image (no specular and sunlight). The specular reflection, direct sunlight, diffuse image, and ambient image can be used for various virtual staging applications\protect\footnotemark[1].}
    \label{fig:framework}
\end{figure*}

%% file: sec1_intro.tex
\section{Introduction}
With the current pandemic-related restrictions and many working from home, there is a significant uptick in home sales. Remote home shopping is becoming more popular and effective tools to facilitate virtual home tours are much needed. One such tool is virtual staging: how would furniture fit in a home and what does it look like if certain settings (\eg sun direction, flooring) are changed? 
To provide effective visualization for staging, panoramas are increasingly being used to showcase homes. Panoramas provide surround information but methods designed for limited-FOV perspective photos cannot be directly applied.

However, inserting virtual furniture into a single panoramic image of a room in a photo-realistic manner is hard. Multiple complex shading effects should be taken into consideration. Such complex interactions are shown in Fig.~\ref{fig:teaser}; the insertion induces effects such as occlusion, sunlight cast on objects, partially shadowed specular reflections, and soft and hard shadows. 
This task is challenging due to the lack of ground truth training data.
The ground truth annotations for appearance decomposition tasks either require significant human labor~\cite{bell2014intrinsic}, or specialized devices in a controlled environment~\cite{grosse2009ground,shen2013real}, which is hard to be extended to a large scale. Previous approaches~\cite{fan2017generic,lin2019deep, li2020inverse} rely on synthetic data for supervised training. However, there is the issue of domain shift and the costs of designing and rendering scenes~\cite{li2020openrooms}.

\footnotetext[1]{Credits: plant\copyright Claudio Ruggio/Adobe Stock, lounge (left)\copyright JaskaranSingh/Adobe Stock, lounge (right)\copyright JaskaranSingh/Adobe Stock, table\copyright Gregory Allen Brown/Adobe Stock, vase\copyright HQ3DMOD/Adobe Stock, glass table\copyright Katerina/Adobe Stock, shelf\copyright Katerina/Adobe Stock, wood texture\copyright nevodka.com/Adobe Stock.}

We present a novel approach to insert virtual objects into a single panoramic image of an empty room in a near-photo-realistic way. Instead of solving a general inverse rendering problem from a single image, we identify and focus on two ubiquitous shading effects that are important for visual realism: interactions of inserted objects with (1) specular reflections and (2) direct sunlight (see Fig.~\ref{fig:teaser}). This is still a challenging problem that has no effective solutions because of the same reasons mentioned above. We observe that while ground truth real data for these effects is hard to obtain, by contrast, it is easier to obtain semantics automatically, given recent advances on layout estimation~\cite{zou2018layoutnet,sun2019horizonnet,pintore2020atlantanet} and semantic segmentation~\cite{long2015fully,chen2017deeplab,wang2020deep}. An automatically generated semantic map with ceiling, wall, floor, window, door, and lamp classes is used to localize the effects coarsely. These coarse localizations are used to supervise a GAN based approach to decompose the input image into four appearance components: (1) diffuse, (2) specular (on floor), (3) direct sunlight (on floor and walls), and (4) diffuse ambient without sunlight. The appearance decompositions can then be used for several applications including insertion of furniture, changing flooring, albedo estimation and estimating and changing sun direction.

We evaluate our method on ZInD~\cite{cruz2021zillow}, a large dataset of panoramas of empty real homes. The homes include a variety of floorings (tile, carpet, wood), diverse configurations of doors, windows, indoor light source types and positions, and outdoor illumination (cloudy, sunny). 
We analyze our approach by comparing against ground truth specular and sunlight locations and sun directions, and by conducting ablation studies. Most previous approaches for diffuse-specular separation and inverse rendering were not designed for the setting used in this work to enable direct apples-to-apples comparisons; they require perspective views or supervised training of large scale real or rendered data. But we nonetheless show performances of such methods as empirical observations and not necessarily to prove that our approach is better in their settings.

Our work also has several limitations: (1) we assume the two shading effects occur either on the planar floor or walls, (2) our methods detect for mid-to-high frequency shading effects but not subtle low-frequency effects or specular/sunlight interreflections, and (3) our methods can induce artifacts if the computed semantic map is erroneous. Despite these limitations, our results suggest that the approach can be an effective and useful tool for virtual staging. Extending the semantics to furniture can enable appearance decomposition of already-furnished homes.

To sum up, our key idea is using easier-to-collect and annotate discrete signals (layout/windows/lamps) to estimate harder-to-collect continuous signals (specular/sunlight). This is a general idea that can be useful for any appearance estimation task. Our contributions include:
(1) A semantically and automatically supervised framework for locating specular and direct sunlight effects (Sec.~\ref{sec:semantic}). (2) A GAN-based appearance separation method for diffuse, specular, ambient, and direct sunlight component estimation (Sec.~\ref{sec:detection} and~\ref{sec:removal}).
(3) Demonstration of multiple virtual staging applications including furniture insertion and changing flooring and sun direction (Sec.~\ref{sec:apps}). To our knowledge, we are the first to estimate direct sunlight and sun direction from indoor images. The overall pipeline is illustrated in Fig.~\ref{fig:framework}.
Our code is released at: \url{https://github.com/tiancheng-zhi/pano_decomp}.

%% file: sec2_related.tex
\section{Related Work}

\paragraph{Inverse Rendering.}
The goal of inverse rendering is to estimate various physical attributes of a scene (\eg geometry, material properties and illumination) given one or more images. Intrinsic image decomposition estimates reflectance and shading layers~\cite{NIPS2002_fa2431bf,li2018learning,janner2017self,li2018cgintrinsics,liu2020unsupervised,baslamisli2021shadingnet}. Other methods attempt to recover scene attributes with simplified assumptions.
Methods~\cite{barron2015, Kim2017ALA,li2018learning,boss2020two} for a single object use priors like depth-normal consistency and shape continuity. Some methods~\cite{Shu2017NeuralFE, Sengupta2018SfSNetLS} use priors of a particular category (e.g., no occlusion for faces). Some works assume near-planar surfaces~\cite{Aittala2015Twoshot,Li_2018_ECCV,hu2022inverse,gao2019deep,deschaintre2019flexible,deschaintre2018single}. In addition, human assistance with calibration or annotation is studied for general scenes~\cite{yu1999inverse, karsch2011}.

Data-driven methods require large amounts of annotated data, usually synthetic images~\cite{li2020inverse,li2021lighting, li2020openrooms,wang2021learning}. The domain gap can be reduced by fine-tuning on real images using self-supervised training via differentiable rendering~\cite{Sengupta2019NeuralIR}. The differentiable rendering can be used to optimize a single object~\cite{kaya2021uncalibrated,zhang2021physg}. Recent works~\cite{boss2021nerd,boss2021neural} extend NeRF~\cite{mildenhall2020nerf} to appearance decomposition. The combination of Bayesian framework and deep networks is explored by ~\cite{chen2021invertible} for reflectance and illumination estimation. In our work, we model complex illumination effects on real $360^\circ$ panoramas of empty homes.
Similar to~\cite{karsch2011}, we believe that discrete semantic elements (like layout, windows, lamps, etc.) are easier to collect and train good models for. By contrast, diffuse and specular annotations are continuous spatially varying signals that are harder to label.

\paragraph{Illumination Estimation.}
Many approaches represent indoor lighting using HDR maps (or its spherical harmonics). Some estimate lighting from a single LDR panorama~\cite{gkitsas2020deep,eilertsen2017hdr}, a perspective image~\cite{gardner2017learning,somanath2020hdr}, a stereo pair~\cite{srinivasan2020lighthouse}, or object appearance~\cite{georgoulis2017reflectance,weber2018learning,park2020seeing}. Recent approaches~\cite{garon2019fast, li2020inverse,wang2021learning} extend this representation to multiple positions, enabling spatially-varying estimation. Others \cite{karsch2014automatic,gardner2019deep,jiddi2020detecting} estimate parametric lighting by modeling the position, shape, and intensity of light sources. Zhang~\etal~\shortcite{zhang2016emptying} combine both representations and estimate a HDR map together with parametric light sources. However, windows are treated as the source of diffuse skylight without considering directional sunlight. We handle the spatially-varying high-frequency sun illumination effects, which is usually a challenging case for most methods. 

\input{fig_noeff}

Some techniques estimate outdoor lighting from outdoor images. Early methods~\cite{lalonde2010sun,lalonde2012estimating} use analytical models to describe the sun and sky. Liu~\etal~\cite{liu2014estimation} estimates sun direction using 3D object models. Recently, deep learning methods~\cite{zhang2019all,hold2019deep,hold2017deep} regress the sun/sky model parameters or outdoor HDR maps by training on large scale datasets. A recent work~\cite{swedish2021objects} estimates high-frequency illumination from shadows. However, they use outdoor images as input, where the occlusion of the sunlight by interior walls is not as significant as that for indoor scenes.

\footnotetext[2]{3D model credits: Same as Fig.~\ref{fig:teaser}.}
\paragraph{Specular Reflection Removal.}
There are two main classes of specular reflection removal techniques. One removes specular highlights on objects. Non-learning based approaches usually exploit appearance or statistical priors to separate specular reflection, including chromaticity-based models~\cite{tan2005separating,yang2010real,shafer1985using,akashi2016separation},
low-rank model~\cite{guo2018single}, and dark channel prior~\cite{kim2013specular}. Recently, data-driven methods \cite{wu2020deep,shi2017learning,fu2021multi} train deep networks in a supervised manner. 
Shi et al.~\shortcite{shi2017learning} train a CNN model using their proposed object-centric synthetic dataset. Fu et al.~\shortcite{fu2021multi} present a large real-world dataset for highlight removal, and introduce a multi-task network to detect and remove specular reflection.
However, the reflection on floors is more complex than highlights, because it may reflect window textures and occupy a large region.

The second class removes reflections from a glass surface in front of the scene. Classical methods use image priors to solve this ill-posed problem, including gradient sparsity~\cite{levin2007user,arvanitopoulos2017single}, smoothness priors~\cite{li2014single,wan2016depth}, and ghosting cues~\cite{shih2015reflection}. Recently, deep learning has been used for this task~\cite{fan2017generic,wan2018crrn,zhang2018single,wei2019single,li2020single,dong2021location,hong2021panoramic} and achieved significant improvements by carefully designing network architectures. 
Li et al.~\shortcite{li2020single} develop an iterative boost convolutional LSTM with a residual reconstruction loss for single image reflection removal. Also, Hong et al.~\shortcite{hong2021panoramic} propose a two-stage network to explicitly address the
content ambiguity for reflection removal in panoramas.
However, they  mostly use supervised training, requiring large amounts of data with ground truth. Most previous works of both classes do not specifically consider the properties of panoramic floor images. 

Our task is in-between these classes, because the reflections are on the floor rather than on glass surfaces but the appearance is similar to that of glass reflection because the floor is flat. To our knowledge, this scenario has not been well studied.

\paragraph{Virtual Staging Services.}
Some companies provide virtual staging services including Styldod\footnote[3]{https://www.styldod.com/}, Stuccco\footnote[4]{https://stuccco.com/}, and PadStyler\footnote[5]{https://www.padstyler.com/}. Users can upload photos of their rooms and the company will furnish them and provide the rendering results. However, most of them are not free and we are not able to know whether the specular reflections and direct sunlight are handled automatically. Certain applications like changing sun direction are usually not supported. By contrast, our approach is released for free use.

%% file: fig_noeff.tex
\begin{figure}
\begin{minipage}{0.74\linewidth}
\includegraphics[width=\linewidth]{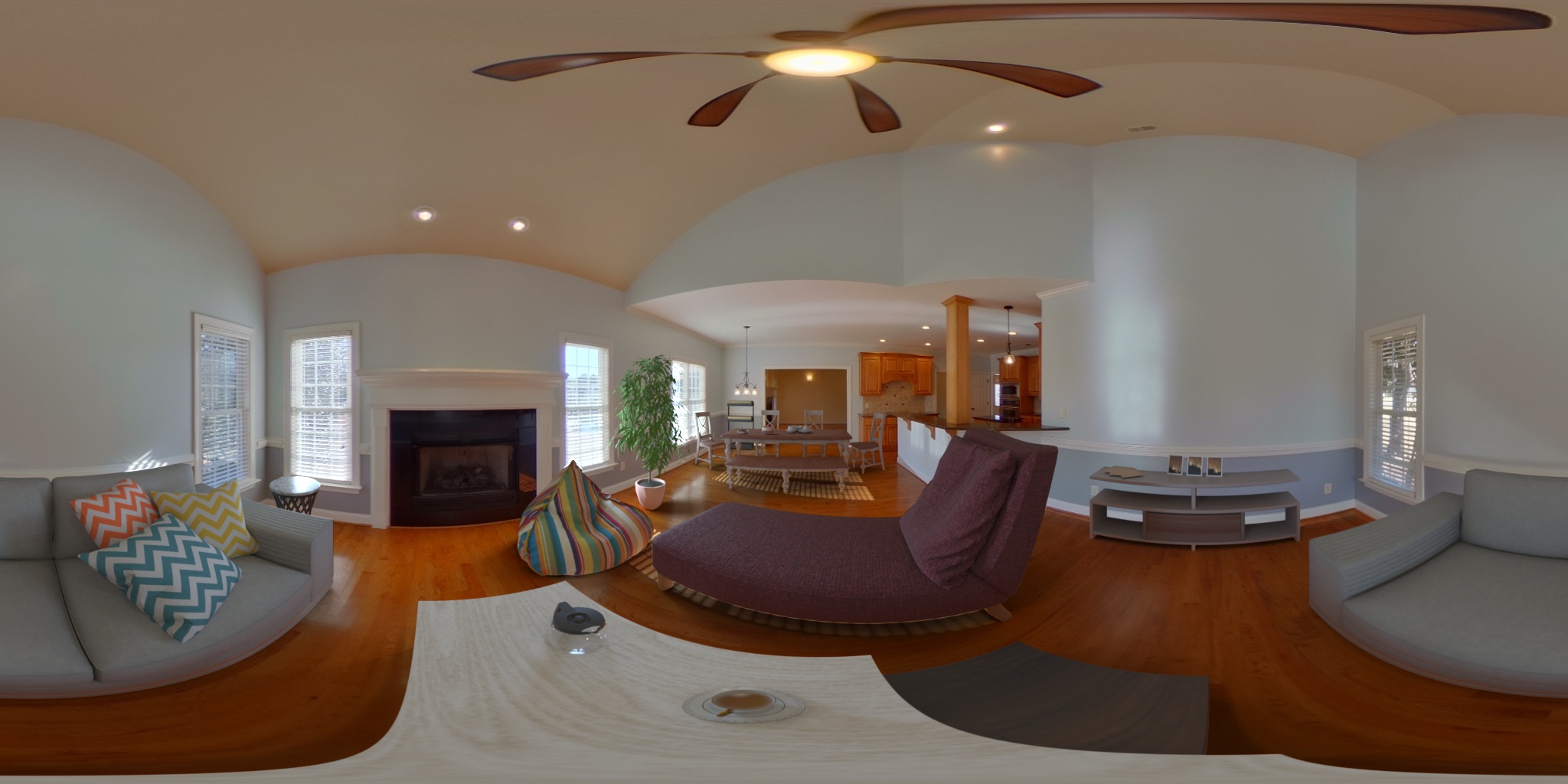}
\end{minipage}
\begin{minipage}{0.244\linewidth}
\includegraphics[width=\linewidth]{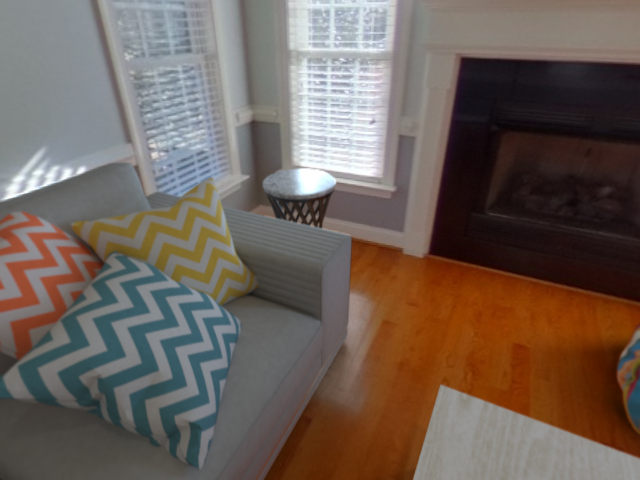} \\
\includegraphics[width=\linewidth]{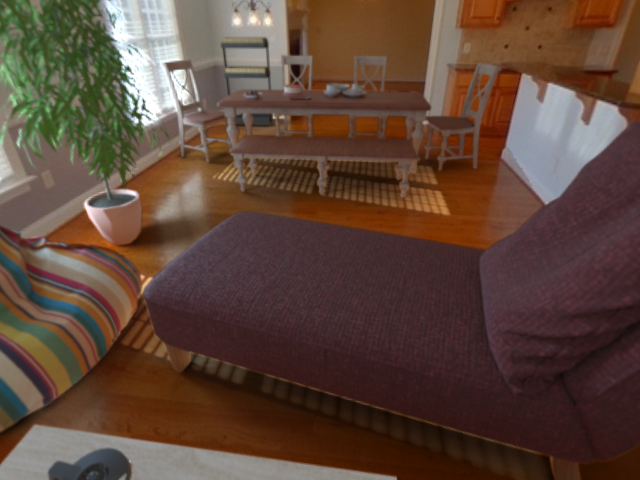}
\end{minipage}
    \caption{Object insertion result\protect\footnotemark[2] without appearance decomposition. Note that the direct sunlight on the objects, hard shadows in the sunlight region, and blocked specular reflections are all missing (compare to Fig.~\ref{fig:teaser}).} 
    \label{fig:noeff}
\end{figure}

%% file: sec4_semantic.tex
\section{Semantics for Coarse Localization of Lighting Effects}
\label{sec:semantic}
\input{fig_geometry}

We focus on two ubiquitous shading effects that are important for virtual staging: (1) specular reflections and (2) direct sunlight. These effects are illustrated in Fig.~\ref{fig:geometry}. Dominant specular reflections in empty indoor environments are somewhat sparse and typically due to sources such as lamps, and open or transparent windows and doors. The sunlight streaming through windows and doors causes bright and often high-frequency shading on the floor and walls. These effects must be located and removed before rendering their interactions with new objects in the environment. However, these effects are hard to annotate in large datasets to provide direct supervision for training. 

Our key observation is that these effects can be localized (at least coarsely) using the semantics of the scene with simple geometrical reasoning. For example, the locations of windows and other light sources (like lamps) constrain where specular reflections occur. The sunlit areas are constrained by the sun direction and locations of windows. Our key insight is that semantic segmentation provides an easier, discrete supervisory signal for which there are already substantial human annotations and good pre-trained models \cite{wang2020deep, MSeg_2020_CVPR}. In this section, we describe how to automatically compute semantic panoramic maps and use them to localize these effects coarsely (Fig.~\ref{fig:coarsemask}(b) and (e)). These coarse estimates are used to supervise a GAN-based approach to refine the locations and extract these effects (Fig. ~\ref{fig:coarsemask} (c) and (f)). 

\input{fig_coarsemask}

\subsection{Transfer Learning for Semantic Segmentation}
\label{ssec:semantic}

We define seven semantic classes that are of interest to our applications: floor, ceiling, wall, window, door, lamp, and other (see Fig.~\ref{fig:coarsemask}(b) for an example). Most works for semantic segmentation are designed for perspective views. Merging perspective semantics to panorama is non-trivial with problems of view selection, inconsistent overlapping regions, and limited FoV. A few networks  ~\cite{su2019kernel,su2017learning} are designed for panoramic images but building and training such networks for our dataset requires annotations and significant engineering effort, which we wish to avoid. Thus, we propose to use a perspective-to-panoramic transfer learning technique, as follows: we obtain an HRNet model~\cite{wang2020deep} pre-trained on a perspective image dataset ADE20K~\cite{zhou2017scene} (with labels adjusted to our setting) and treat it as the ``Teacher Model''. Then we use the same model and weights to initialize a ``Student Model'', and adapt it for panoramic image segmentation.

To supervise the Student Model, we sample perspective views from the panorama. Let $I_{pano}$ be the original panorama, $\Phi$ be the sampling operator, $f_{tea}$ be the Teacher Model function, and $f_{stu}$ be the Student Model function. The transfer learning loss $L_{trans}$ is defined as the cross entropy loss between $f_{tea}(\Phi(I_{pano}))$ and $\Phi(f_{stu}(I_{pano}))$. To regularize the training, we prevent the Student Model from deviating from the Teacher Model too much by adding a term $L_{reg}$ defined as the cross entropy loss between $f_{tea}(I_{pano})$ and $f_{stu}(I_{pano})$. The total loss is $L=w_{trans} L_{trans} + w_{reg} L_{reg}$. $w_{trans}$ and $w_{reg}$ are weights given by the confidence of the Teacher Model prediction $f_{tea}(\Phi(I_{pano}))$ and $f_{tea}(I_{pano})$.

The semantic map is further improved via HDR map and layout estimation. See Sec.~\ref{ssec:implement} for details. Fig.~\ref{fig:coarsemask}(b) shows an example semantic map computed using our approach. 

\subsection{Coarse Localization of Lighting Effects}
\label{ssec:coarsemask}

\paragraph{Coarse Specular Mask Generation.}
The panoramic geometry dictates that \textit{the possible specular area, on the floor, is in the same columns as the light sources}. This assumes that the camera is upright and that the floor is a mirror. See Fig.~\ref{fig:geometry} for an illustration. Thus, we simply treat floor pixels, that are in the same columns as the light sources (windows and lamps in the semantic map), as the coarse specular mask. To handle rough floors, we dilate the mask by a few pixels. See Fig.~\ref{fig:coarsemask}(c) for an example coarse specular mask.

\paragraph{Coarse Sunlight Mask Generation.}
We first estimate a coarse sun direction and then project the window mask to the floor and walls according to this direction. 
Since sunlight enters the room through windows (or open doors), the floor texture can be projected back onto the window region based on the sun direction. This projected floor area must be bright and a score is estimated as the intensity difference between window region and projected floor. The projection requires room geometry which can be obtained by calculating wall-floor boundary from the semantic map. Basically, we assume an upright camera at height=1 and the floor plane at Z=-1. Then, the 3D location (XYZ) of a wall boundary point is computed using known $\theta-\phi$ from the panorama, Z=-1, and cartesian-to-spherical coordinate transform, which can be used to calculate the wall planes.

For a conservative estimate of the floor mask, we use the K-best sun directions. Specifically, the sun direction is given by $\mathbf{d}=(\theta,\phi)$, where $\theta$ is the elevation angle and $\phi$ is the azimuth angle. We do a brute-force search for $\theta$ and $\phi$ with step size $=5^\circ$. Let $s(\theta,\phi)$ be the matching score. First, top-K elevation angles $\theta_1, ..., \theta_K$ are selected based on score $s'(\theta)=\max_\phi s(\theta,\phi)$. Let $\phi^*(\theta)=\argmax_\phi s(\theta,\phi)$ be the optimal azimuth angle for elevation $\theta$. Then the selected sun directions are $(\theta_1, \phi^*(\theta_1)),...,(\theta_K, \phi^*(\theta_K))$. In practice, we set $K=5$. Fig.~\ref{fig:coarsemask} (e) for an example coarse direct sunlight mask.

%% file: fig_geometry.tex
\begin{figure}
    \centering
    \subfloat[(a) Specular Reflection]{
    \includegraphics[width=0.49\linewidth]{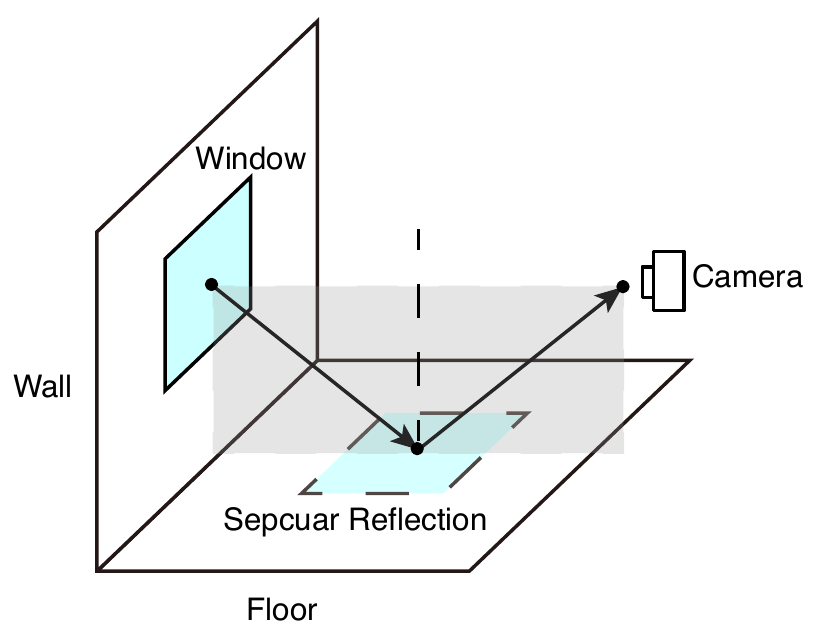}}
    \subfloat[(b) Direct Sunlight]{
    \includegraphics[width=0.49\linewidth]{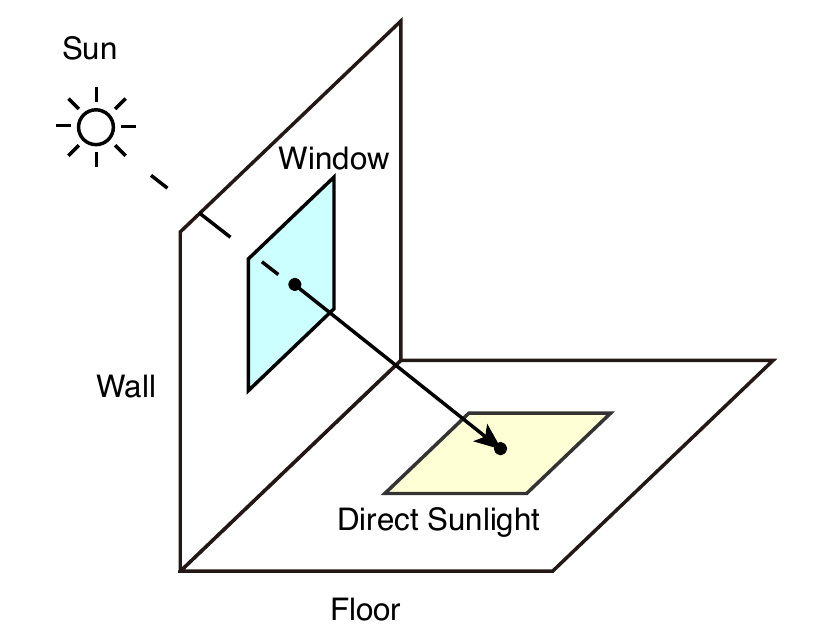}}
    \caption{The geometry of lighting effects considered in this work. (a) For specular reflections, when the camera is upright, the light source point (transparent or open window, door, or indoor lamp), the reflection point, and the camera center lie on a vertical plane (shown as gray), corresponding to a column in the panorama. (b) For direct sunlight, the sun direction establishes the mapping between a window (or door) point and a floor point illuminated by direct sunlight. The sunlit floor area can be back-projected to the window according to the sun direction. }
    \label{fig:geometry}
\end{figure}

%% file: fig_coarsemask.tex
\begin{figure}
    \centering
    \includegraphics[width=0.7\linewidth]{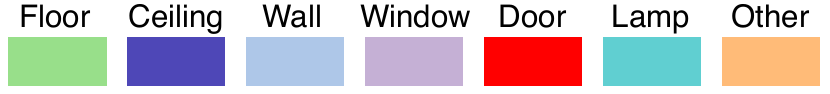}
    
    \subfloat[(a) RGB]{
    \includegraphics[width=0.325\linewidth]{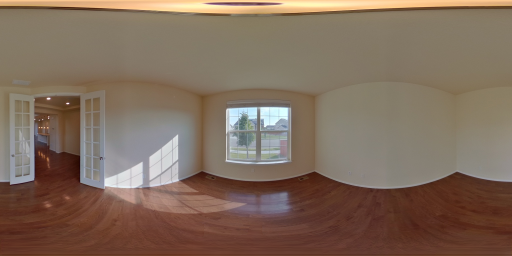}}
    \subfloat[(b) Coarse Specular Mask]{
    \includegraphics[width=0.325\linewidth]{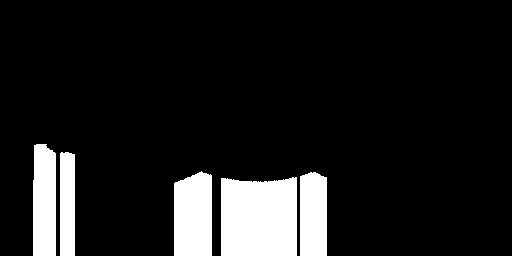}}
    \subfloat[(c) Fine Specular Image]{
    \includegraphics[width=0.325\linewidth]{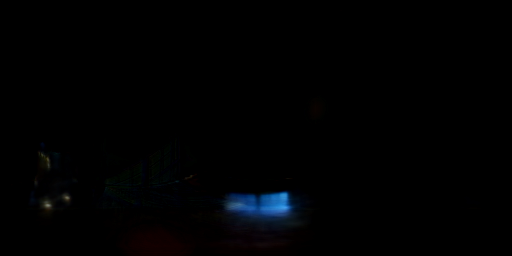}}

    \subfloat[(d) Semantic Map]{
    \includegraphics[width=0.325\linewidth]{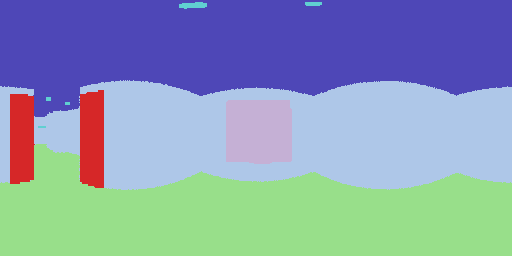}}
    \subfloat[(e) Coarse Sunlight Mask]{
    \includegraphics[width=0.325\linewidth]{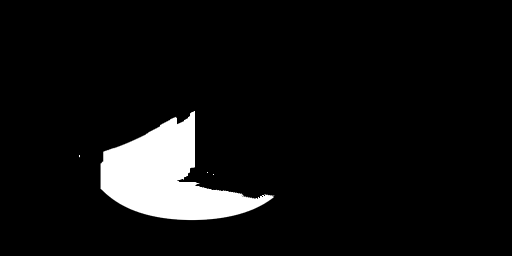}}
    \subfloat[(f) Fine Sunlight Image]{
    \includegraphics[width=0.325\linewidth]{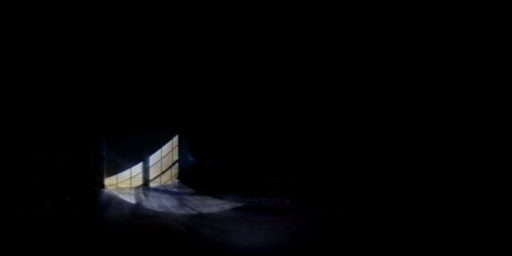}}

    \caption{A semantic map with 7 classes is computed automatically from a panorama. The map is used to obtain coarse lighting effect masks exploiting the geometric constraints in Fig.~\ref{fig:geometry}. These coarse masks are used to supervise a GAN-based method to obtain accurate specular and sunlit images.}
    \label{fig:coarsemask}
\end{figure}

%% file: sec5_detection.tex
\section{GAN-Based Lighting Effects Detection}
\label{sec:detection}
Both the coarse masks estimated above are rough conservative estimates. In this section, we will demonstrate that this inaccurate information can still be used as a supervision signal that can effectively constrain and guide an accurate estimation of specular reflection and sunlit regions. We sketch the key idea in Sec.~\ref{ssec:idea}, and explain the detailed loss function in Sec~\ref{ssec:loss}.

\input{fig_architecture}

\subsection{Key Idea and Overall Architecture}
\label{ssec:idea}
Simply training a lighting effects network with the coarse masks will result in the network predicting only the coarse masks. How do we design a network that is able to produce more accurate masks? Fig.~\ref{fig:architecture} illustrates our architecture that augments the lighting effects network with three local discriminators. The lighting effects network takes a single panoramic image $I$ as an input, and predicts the specular ($I_{spec}$) and sunlight ($I_{sunl}$) components. The first local discriminator takes the image with specular component removed ($I-I_{spec}$) as an input and tries to locate the specular reflection via per-pixel binary classification. Its output is denoted by $D_{spec}$. It is supervised by a coarse specular mask $M_{spec}$ obtained from semantics described in Sec.~\ref{ssec:coarsemask}. If the specular prediction $I_{spec}$ is good, the discriminator should fail to locate specular reflection on image $I-I_{spec}$. Hence, the lighting effects networks should try to fool the discriminator. This idea can be seen as a per-pixel GAN treating the pixels inside the coarse mask as fake examples and the pixels outside the mask as real examples. This approach is also applied to sunlight regions, where the sunlight discriminator output is denoted by $D_{sunl}$ and the coarse sunlight mask is $M_{sunl}$.


Because artifacts could appear when specular reflection and direct sunlight overlap with each other if the intensities are not predicted accurately, we include a third discriminator trained to detect overlapping specular reflection and direct sunlight on the image with both effects removed ($I-I_{spec}-I_{sunl}$). Its prediction is denoted by $D_{over}$. This discriminator is supervised by the intersection of the coarse specular mask and the coarse sunlight mask, represented as $M_{spec}\odot M_{sunl}$, where $\odot$ is the element-wise product.

\subsection{Loss Function}
\label{ssec:loss}

\paragraph{Sparsity Loss.}
We observe that, in an empty room, specular reflection usually appears on the floor, and direct sunlight usually appears on the floor and the walls. Thus, we define the region of interest (ROI) for specular reflection to be the floor and the ROI for direct sunlight to be the floor and the walls. Different regions should have different levels of sparsity.
Lighting effects should be sparser outside the ROI than inside the ROI. They should also be sparser outside the coarse mask than inside the coarse mask.
Thus, we apply different levels of sparsity to different regions. Let $I$ be the input image, $R$ be the binary ROI mask, and $M$ be the coarse mask inside the ROI. The sparsity loss is:
\begin{equation}
L_{s} = \lambda_{s1} \lVert I \odot R \rVert_1 + \lambda_{s2} \lVert I \odot R \odot (1-M) \rVert_1 + \lambda_{s3} \lVert I \odot M \rVert_1, 
\end{equation}
where, $\lambda_{s1} \geq \lambda_{s2} \geq \lambda_{s3}$ are constant weights, and $\odot$ is the element-wise product.
We use such sparsity loss for specular reflection and direct sunlight separately, denoted by $L_{sp\_s}$ and $L_{sl\_s}$.

\paragraph{Adversarial Loss.}
As mentioned in Sec.~\ref{ssec:idea}, after removing the lighting effects, a discriminator should not be able to locate the effects. Consider specular reflection as an example. Let $I$ be the input image, and $I_{spec}$ be the estimated specular reflection. We use a local discriminator to find reflection areas from $I - I_{spec}$. The discriminator is supervised by coarse specular mask $M_{spec}$. The discriminator loss uses pixel-wise binary cross entropy, $L_{sp\_d}= f_{ce}(D_{spec}, M_{spec}) $. To fool the discriminator, an adversarial loss:
\begin{equation}
    L_{sp\_a}= f_{ce}(D_{spec}, 1-M_{spec})
\end{equation}
is applied to region $M_{spec}$.
Similarly, we build adversarial losses for sunlight regions and regions with overlapping specular reflection and direct sunlight, denoted by $L_{sl\_a}$ and $L_{ol\_a}$, respectively.

\paragraph{Total Loss.}
The total loss $L$ is the sum of the two sparsity losses and the three adversarial losses:
\begin{equation}
    L = \lambda_{sp\_s}L_{sp\_s} + \lambda_{sl\_s}L _{sl\_s} + \lambda_{sp\_a}L_{sp\_a} + \lambda_{sl\_a}L_{sl\_a} + \lambda_{ol\_a}L_{ol\_a},
\end{equation}
where $\lambda_{sp\_s}$, $\lambda_{sl\_s}$,$\lambda_{sp\_a}$,$\lambda_{sl\_a}$, $\lambda_{ol\_a}$ are constant weights. 

%% file: fig_architecture.tex
\begin{figure}
    \centering
    \includegraphics[width=\linewidth]{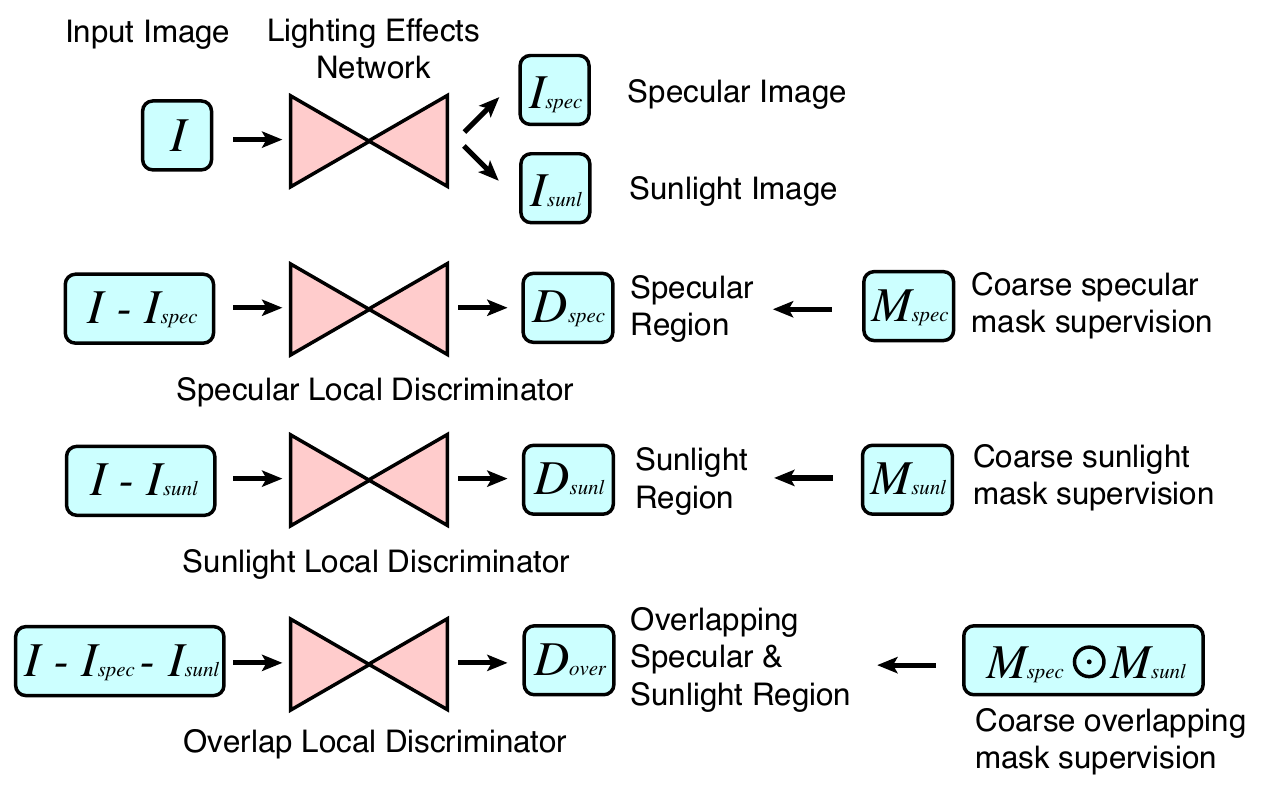}
    \caption{Architecture for lighting effects detection. The lighting effects network takes a panorama $I$ as input, and predicts the specular ($I_{spec}$) and sunlight ($I_{sunl}$) components. If the prediction is good, a local discriminator trying to locate specular regions should fail on the image with specular component removed ($I-I_{spec}$). This is the key supervision signal for training the lighting effects network. The local discriminator is supervised by coarse specular masks obtained from semantics. Similar techniques are applied to sunlight regions and regions with overlapping specular and direct sunlight.}
    \label{fig:architecture}
\end{figure}

%% file: sec6_removal.tex
\section{Lighting Effects Removal}
\label{sec:removal}
\input{fig_remove_mot}
The previous architecture and method focused on estimating the specular and direct sunlight regions. But, naively subtracting the predicted regions from the original image, i.e. $I - I_{spec}$, $I - I_{spec}- I_{sunl}$, may still produce artifacts as shown in Fig.~\ref{fig:remove_motiv}. In this section, we instead directly estimate the diffuse image (without specular reflection) and the ambient image (without specular reflection or direct sunlight).

\subsection{GAN-Based Specular Reflection Removal}
\label{ssec:weak}
\input{fig_diffarch}
As shown in Fig.~\ref{fig:diffarch}, a deep network is used to predict a diffuse image. To supervise the training of the network, we adopt the same GAN strategy as Sec.~\ref{ssec:loss}. Instead of using the coarse mask generated by Sec.~\ref{ssec:coarsemask}, we obtain a fine specular mask by thresholding the estimated specular reflection for discriminator supervision.

Let $I$ be the original image, $I_{spec}$ be the specular image, $I_{diff}$ be the diffuse image, and $I_{recon} = I_{diff} + I_{spec}$ be the reconstructed image. Inspired by \cite{zhang2018single}, our loss $L_{diff}$ consists of an adversarial loss $L_{adv}$, a reconstruction loss $L_{recon}$, a perceptual loss $L_{perc}$, and an exclusion loss $L_{excl}$:
\begin{equation}
L_{diff} = \lambda_a L_{adv} + \lambda_r L_{recon} + \lambda_p L_{perc} + \lambda_e L_{excl},
\end{equation}
where $\lambda_a$, $\lambda_r$, $\lambda_p$,  $\lambda_e$ are constant weights. 

\paragraph{Adversarial Loss.} $L_{adv}$ is the same as the loss defined in Sec.~\ref{ssec:loss}, replacing the coarse specular mask by the fine specular mask.

\paragraph{Reconstruction Loss.}
$L_{recon} =\lVert I_{recon} - I \rVert_1$ calculates the L1 loss between the reconstructed image and the original image.

\paragraph{Perceptual Loss.} Following~\cite{johnson2016perceptual} and \cite{zhang2018single}, $L_{perc} = \lVert\Phi(I_{recon})-\Phi(I)\rVert_1$ calculates the L1 distance between VGG features~\cite{simonyan2014very}, where $\Phi$ is the VGG feature function. It is for enhancing perceptual quality.

\paragraph{Exclusion Loss.} $L_{excl} = \lVert tanh (|\nabla I_{diff}|) \odot tanh(|\nabla I_{spec}|) \rVert_F$, where $\lVert \cdot \rVert_F$ is the Frobenius norm, and $\odot$ is the element-wise product. This loss minimizes the correlation between diffuse and specular components in the gradient domain to prevent edges from appearing in both diffuse and specular images, following~\cite{zhang2018single}.

\subsection{Semantic-Aware Inpainting for Sunlight Removal}
A process similar to that used for diffuse image estimation can be used for predicting the ambient image without sunlight. In practice, however, bright sunlit regions are the highly dominant brightness component and are often saturated and the network does not predict the ambient image well. Thus, we use an inpainting approach called SESAME~\cite{ntavelis2020sesame} that is semantic-aware and preserved the boundaries between different classes (wall versus floor, etc).

\label{ssec:strong}

%% file: fig_remove_mot.tex
\begin{figure}
    \centering
    \subfloat[{(a) Naive Diffuse}]{
    \includegraphics[width=0.49\linewidth]{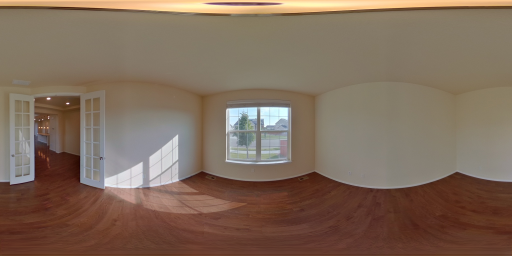}}
    \subfloat[{(b) Naive Ambient}]{
    \includegraphics[width=0.49\linewidth]{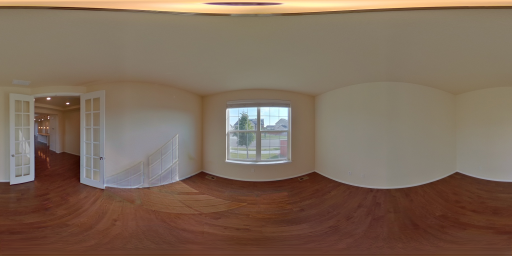}}

    \subfloat[{(c) Our Diffuse}]{
    \includegraphics[width=0.49\linewidth]{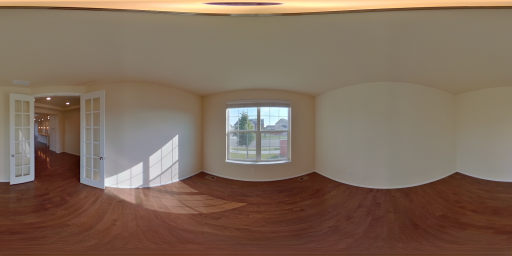}}
    \subfloat[{(d) Our Ambient}]{
    \includegraphics[width=0.49\linewidth]{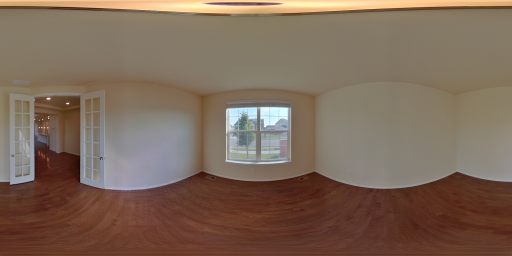}}

    \subfloat[(e) Inpainting Mask]{
    \includegraphics[width=0.49\linewidth]{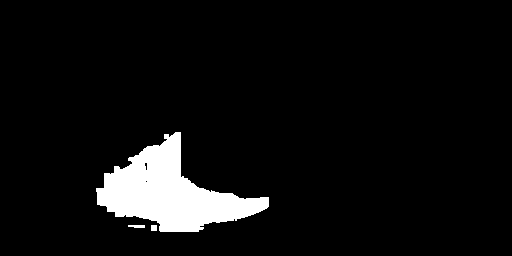}}
    \subfloat[{(f) Brightened Samples}]{
    \includegraphics[width=0.49\linewidth]{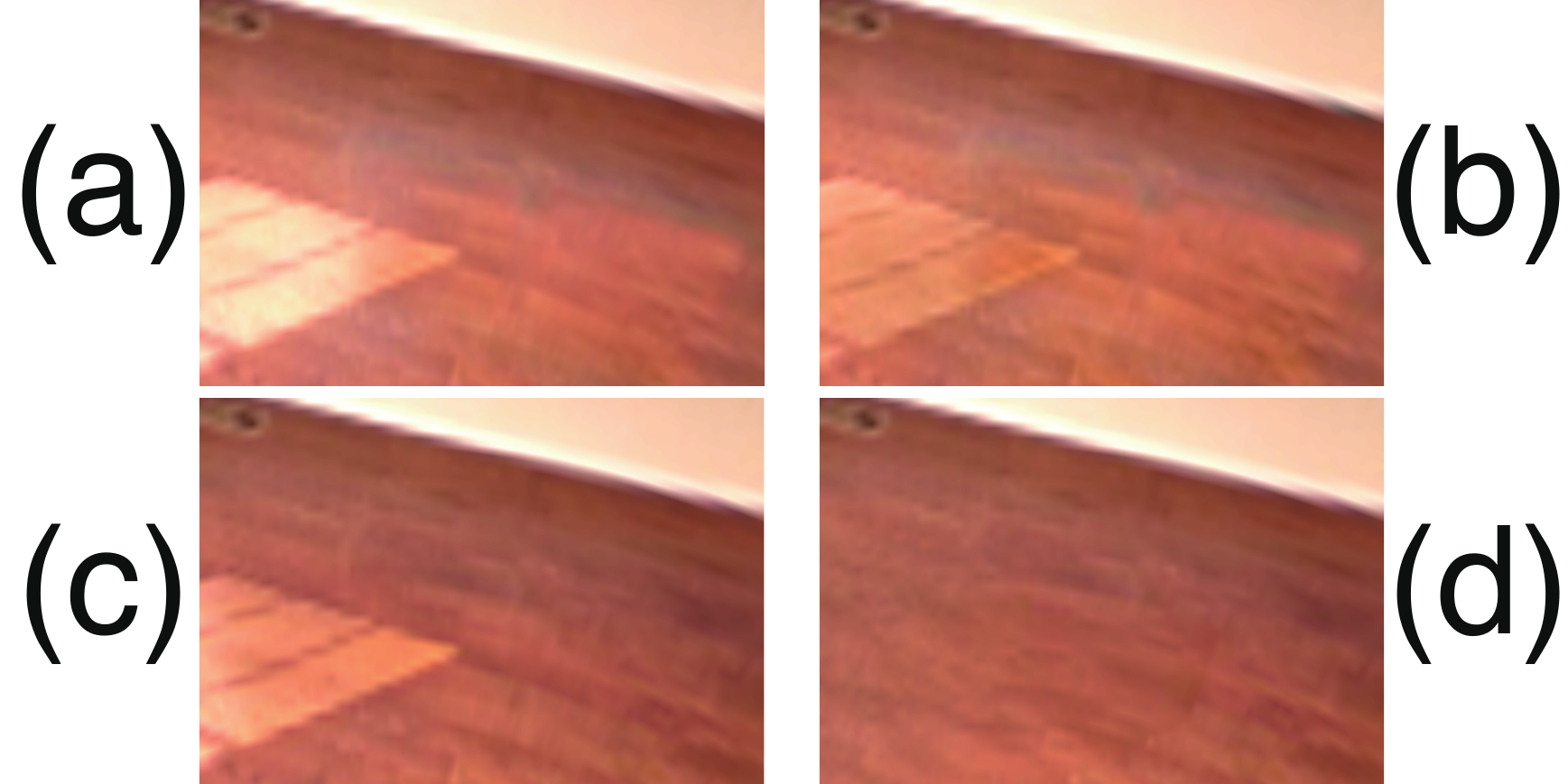}}
    \caption{Removing effects by naively subtracting the estimated lighting effects versus our approach. Naive subtraction ($I-I_{spec}$, $I-I_{spec}-I_{sunl}$) leaves lighting effects residues (specular/sunlight region is brighter than the surrounding, with visible boundaries). Our result removes these artifacts.}
    \label{fig:remove_motiv}
\end{figure}

%% file: fig_diffarch.tex
\begin{figure}
    \centering
    \includegraphics[width=\linewidth]{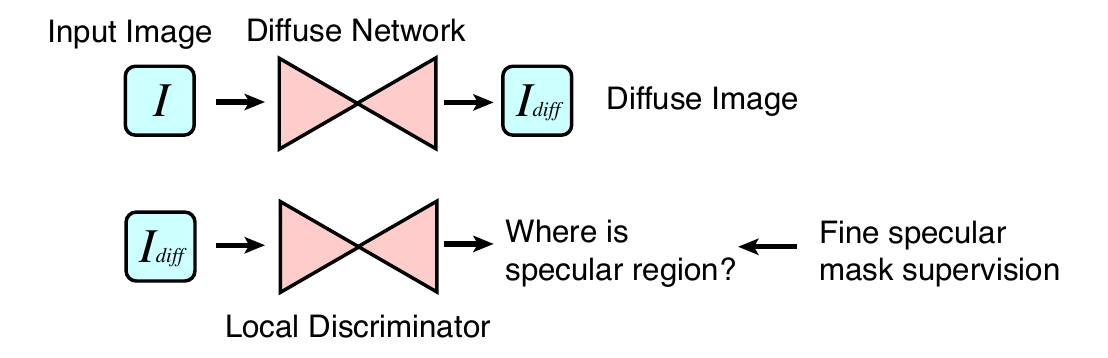}
    \caption{Architecture for specular reflection removal. The diffuse network takes a single panoramic image $I$ as input, and predicts the diffuse component $I_{diff}$. Similar to Fig.~\ref{fig:architecture}, a local discriminator trying to locate specular regions should fail on the diffuse image $I_{diff}$. The discriminator is supervised by the fine specular mask obtained via thresholding and dilating the estimated specular component.}
    \label{fig:diffarch}
\end{figure}

%% file: sec7_eval.tex
\section{Experimental Analysis}

\subsection{Implementation Details}
\label{ssec:implement}

\paragraph{Improving Lamp Semantics with HDR Estimation.}
The ADE20K definition of lamps includes the whole lamp, while we care about the bright bulb. Thus, we estimate a HDR map via supervised learning to obtain a more accurate bulb segmentation. To obtain a HDR image $I_{hdr}$, we train U-Net~\cite{ronneberger2015u} to predict the HDR-LDR residue in log space $I_{res}=ln(1+I_{hdr}-I_{ldr})$, using Laval~\cite{gardner2017learning} and HDRI Heaven~\footnote[6]{https://hdrihaven.com/hdris/?c=indoor} datasets. A weighted L2 loss is adopted. We set batch size 32, learning rate $10^{-4}$ and epochs 300. Pixels in the lamp segmentation with $I_{hdr} > 2$ are kept.

\paragraph{Improving Room Semantics with Layout Estimation.}
The floor-wall and ceiling-wall boundaries predicted by semantic segmentation might not be straight in perspective views, causing visual artifacts. Thus, we adopt a layout estimation method LED$^2$-Net~\cite{Wang_2021_CVPR} to augment the semantic map. The layout estimation predicts floor, wall and ceiling regions for the current room, which is used to improve the floor-wall and ceiling-wall boundaries. We adopt batch size 6, learning rate $3\times10^{-4}$, and epochs 14.

\paragraph{Super-Resolution in Post-Processing.}
Due to limited computing, we run our methods at low-resolution ($256 \times 512$). We enhance the resolution for applications requiring high resolution results with post-processing.  We enhance the resolution of the specular image to $1024 \times 2048$ via a pre-trained super-resolution model~\cite{wang2021real}. To ensure the consistency between the diffuse image and the original image, we adopt Deep Guided Decoder~\cite{uezato2020guided} for super-resolution with the original image as guidance. The final resolution is $1024 \times 2048$. For the ambient image, we simply use super-resolution~\cite{wang2021real} and copy-paste the inpainted region on to the high resolution diffuse image. To preserve high-frequency details, we calculate the difference between the diffuse and ambient components as the high-resolution sunlight image.

\paragraph{Deep Architecture and Hyper-parameters.}
For semantic transfer learning, we use $w_{trans}=w_{reg}=1$, batch size 4, learning rate $10^{-5}$, epochs 5, and random perspective crops with FoV 80$^\circ$ to 120$^\circ$, elevation -60$^\circ$ to 60$^\circ$, azimuth 0$^\circ$ to 360$^\circ$.
We use U-Net~\cite{ronneberger2015u} for Lighting Effects Network and Diffuse Network, and FCN~\cite{long2015fully} for discriminators, optimized with Adam~\cite{kingma2014adam}. See supplementary materials for network architectures.
The networks use learning rate $10^{-4}$, weight decay $10^{-5}$, and batch size 16.
For effects detection, $\lambda_{sp\_s}=\lambda_{sl\_s}=\lambda_{sp\_a}=\lambda_{sl\_a}=\lambda_{ol\_a}=1$. For specular sparsity, $\lambda_{s1}=10, \lambda_{s2}=3, \lambda_{s3}=1$. For sunlight sparsity, $\lambda_{s1}=5, \lambda_{s2}=\lambda_{s3}=1$. For specular reflection removal, $\lambda_{r}=\lambda_{p}=\lambda_{e}=1, \lambda_{a}=10$. The networks take $\sim$2 days for training on 4 TITAN Xp GPUs and 0.05s for inference. Resolution enhancement using Guided Deep Decoder~\cite{uezato2020guided} converges in $\sim$30 min for a single image.

\paragraph{Dataset.}
We evaluate our approach on the ZInD~\cite{cruz2021zillow} dataset that includes 54,034 panoramas for training and 13,414 panoramas for testing (we merge their validation and test sets).
For better visualization, we inpaint the tripod using SESAME~\cite{ntavelis2020sesame}.
Sec.~\ref{ssec:ablation} and \ref{ssec:compare} evaluate our approach at $256\times 512$ resolution. Sec.~\ref{ssec:eval_hr} visualizes results at resolution $1024\times 2048$.

\subsection{Ablation Study}
\label{ssec:ablation}
\paragraph{Perspective Merging vs. Panoramic Semantics.} We compare with merging perspective semantics to panorama, by sampling 14 perspective views and merging the semantics via per-pixel voting. On Structured3D~\cite{zheng2020structured3d} test set (mapped to our labels), the merging method achieves mIoU=44.8\%, while ours achieves 68.1\%. This shows developing methods directly is meaningful. We also test our method on different emptiness levels: No furniture mIoU: 68.1\%; Some furniture: 66.6\%; Full furniture: 60.6\%. Since ZInD does not contain much furniture, this mIoU reduction is expected.

\paragraph{Lighting Effects Detection.}
We manually annotate specular and sunlight regions on 1,000 test panoramas sampled from the test set for quantitative analysis. The annotation is coarse because there is no clear boundary of specular regions. For a fair comparison, we report the mean IoU with respect to the best threshold for binarization for each method. The quantitative evaluation is performed on the floor region only.

To show that the GAN-based method introduces priors for refining the coarse mask, we train a network with the same architecture as Lighting Effects Network in a standard supervised manner (without GAN), using the coarse mask as labels. We also compare with the coarse mask itself and the method without the third discriminator for overlap regions. Tab.~\ref{tab:eff_ablation} shows that the full method outperforms the others, demonstrating the effectiveness of our GAN-based approach with three discrimintors.

\paragraph{Specular Reflection Removal.}
We use synthetic data for evaluating specular reflection removal. We render 1,000 images based on microfaucet reflectance model using the Mitsuba~\cite{Mitsuba} renderer with floor textures from AdobeStock. We also define images with peak specular intensity ranking top 5\% in the test data as ``strong reflection" . PSNR (higher is better) and LPIPS~\cite{zhang2018unreasonable} (lower is better) are reported.

We compare with the naive method by subtracting the specular component from the original image directly, and methods without each loss components. In Tab.~\ref{tab:diff_ablation}, the full method outperforms the other variants on all metrics except for the ones without adversarial loss. Although the methods without adversarial loss achieves a high score on ``all testdata", it performs worse than the full method on ``strong reflection". Besides, as shown in Fig.~\ref{fig:loss}, adversarial and exclusion losses help remove perceivable visual artifacts or specular residues although average metrics don’t reflect clearly. Thus, we adopt the full method with all four losses for the applications.

\input{tab_eff_ablation}
\input{tab_diff_ablation}

\subsection{Performances of Other Relevant Approaches}
\label{ssec:compare}
Most previous approaches for diffuse-specular separation and inverse rendering were not designed for the setting used in this work to enable direct apples-to-apples comparisons; they require perspective views or supervised (ground truth) training of large scale real or rendered data or are used for a related task. 
But accurately annotating large data for specular and sunlight effects is non-trivial, because of blurred boundaries or complex texture. The visual and quantitative evaluations below shows that previous approaches do not easily bridge the domain gap. These methods have strong value for the domains they were designed for but our approach provides an effective tool for virtual staging with real-world panoramas.

\paragraph{Specular Reflection Detection.}
We evaluate a bilateral filtering based method BF ~\cite{yang2010real}, two deep reflection removal methods IBCLN~\cite{li2020single} and LASIRR~\cite{dong2021location}, and a deep highlight removal method JSHDR~\cite{fu2021multi}. 
Since these methods are designed for perspective images, we randomly sample 5,000 perspective crops from the 1,000 annotated panoramic images and report the performance in Tab.~\ref{tab:spec_compare} left column. In the right column, we also evaluate in panoramic domain by converting the panorama to a cubemap, running the baseline methods, and merging the results back to panorama. Among the evaluated relevant methods, JSHDR~\cite{fu2021multi} achieves the best performance, likely because it is designed for specular highlight removal, which is closer to our setting than glass reflection removal methods.

\input{fig_loss}
\input{fig_compspec}

\input{tab_spec_compare}

\input{tab_diff_compare}
\input{fig_highres}

\paragraph{Direct Sunlight Detection.}
We are unaware of any work estimating direct sunlight for our setting. Thus, we evaluate an intrinsic image decomposition method USI3D~\cite{liu2020unsupervised}. We assume that the sunlight region should have a strong shading intensity and threshold the output shading image as the sunlight prediction. This achieves mean IoU 7.4\% on perspective crops and 7.9\% on panoramas from cubemap merging while our method achieves 47.2\% on perspective crops and 47.2\% on panoramas.  We conclude that using the semantics and sun direction is important for sunlight estimation.

\paragraph{Specular Reflection Removal.}
Similar to lighting effect detection, we report performances based on 5,000 perspective crops sampled from the 1,000 synthetic panoramic images. Fig.~\ref{fig:compspec} and the first four rows in Tab.~\ref{tab:diff_compare} show the performances of pre-trained models on perspective images. Similar to specular reflection detection, JSHDR~\cite{fu2021multi} achieves the best performance, likely due to smaller domain differences, since it is trained on a large-scale real dataset. A similar conclusion can be drawn from the evaluation in panoramic domain.
We also attempted to bridge the domain gap for a more reasonable quantitative evaluation. For JSHDR, only the compiled code is available online, making it hard to adapt to our domain. Thus, we train IBCLN on our panoramic data. Specifically, let $\{I\}$ be the set of original images and $\{I_{spec}\}$ be the set of estimated specular components. We treat $\{I-I_{spec}\}$ and $\{I_{spec}\}$ as the transmission image set and the reflection image set, respectively. These two sets are then used for rendering the synthetic images for training. Tab.~\ref{tab:diff_compare} Rows 5 and 11 show that by re-training using the synthetic data based on our specular reflection estimation, the performance of IBCLN improves but still does not reach the level of our method.

\subsection{Diverse Lighting and High Resolution Results}
\label{ssec:eval_hr}
Fig.~\ref{fig:special} visualizes effects detection results in diverse lighting conditions, including large sunlight area and (partially) occluded light sources. The results show that our detection can handle such cases.

\input{fig_special}

Fig.~\ref{fig:highres} visualizes high-resolution appearance decomposition results. The specular reflections are well removed and the floor texture is consistent with the original image. The sunlight region is inpainted with the correct colors. Due to the limitation of the inpainting algorithm, the texture is not perfect. As inpainting algorithms improve, we can plug-and-play those within our pipeline.

%% file: tab_eff_ablation.tex
\begin{table}[]
    \centering
    \caption{Ablation study for lighting effects detection on real data. mIoU (\%) for specular and sunlight regions are reported. Our full method performs better than the others, showing the effectiveness of the GAN-based design.}
    \scalebox{0.805}{
    \begin{tabular}{lccc}\toprule
    Method & Specular $\uparrow$ & Sunlight $\uparrow$ \\
    \midrule
    Coarse mask & 6.7 & 3.0 \\
    Supervised by coarse mask w/o GAN & 8.7 & 19.9\\
    No overlap effects discriminator & 34.6 & 30.0\\
    Our full method  & \textbf{38.9} & \textbf{47.2} \\
    \bottomrule
    \end{tabular}}
    \label{tab:eff_ablation}
\end{table}

%% file: tab_diff_ablation.tex
\begin{table}[]
    \centering
    \caption{Ablation study for specular removal on synthetic panoramas. Note that specular reflections are not strongly visible in all images of the dataset and they are sparse when visible. Hence, we also report the metrics on a subset of images with strong reflections. Although "no adversarial loss" performs better on "all testdata", it does not reach the level of the full method on strong reflections, which significantly affect the visual quality.}
    \scalebox{0.805}{
    \begin{tabular}{lccccc}\toprule
    \multirow{2}*{Method}  & \multicolumn{2}{c}{All Testdata} & \multicolumn{2}{c}{Strong Reflection}  \\ \cmidrule{2-5}& PSNR $\uparrow$ & LPIPS $\downarrow$ &PSNR $\uparrow$&LPIPS $\downarrow$\\
    \midrule
    Input subtracts specular image & 33.2 & 0.0247 & 25.0 & 0.0702\\
    No adversarial loss  &\textbf{33.4}& 0.0233 & 25.2 & 0.0681\\
    No reconstruction loss &33.2&0.0248 & 25.1 & 0.0693\\
    No perceptual loss &31.3&0.0483 & 25.0 &0.0903 \\
    No exclusion loss &33.3&0.0242&25.2&0.0693\\
    No adversarial/reconstruction &\textbf{33.4} &\textbf{0.0232}&25.1&0.0683 \\
     No adversarial/perceptual &\textbf{33.4} &0.0270 & \textbf{25.3} & 0.0749\\
    No adversarial/exclusion &33.3 &0.0233 & 25.1& 0.0684 \\
    No reconstruction/perceptual & 14.4 & 0.3595 & 12.7 & 0.3407\\
    No reconstruction/exclusion & 33.2 & 0.0242 & 25.2 & 0.0700\\
    No perceptual/exclusion& 32.4 & 0.0333 & 25.1 & 0.0800\\
    Our full method &33.3  & 0.0241 & \textbf{25.3} & \textbf{0.0676} \\
    \bottomrule
    \end{tabular}}
\label{tab:diff_ablation}
\end{table}

%% file: fig_loss.tex
\begin{figure}
    \centering
\subfloat[(a) RGB]{
\includegraphics[width=0.325\linewidth]{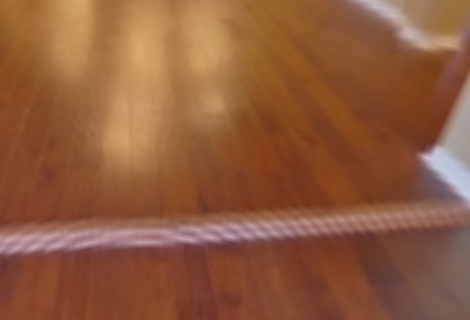}}
\subfloat[(b) No Adversarial]{
\includegraphics[width=0.325\linewidth]{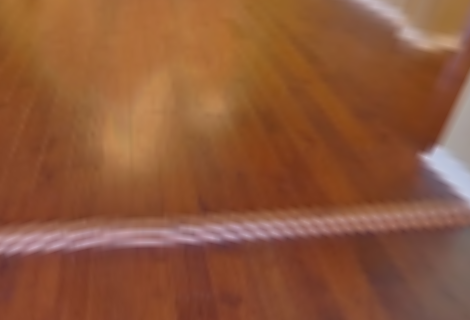}}
\subfloat[(c) No Reconstruction]{
\includegraphics[width=0.325\linewidth]{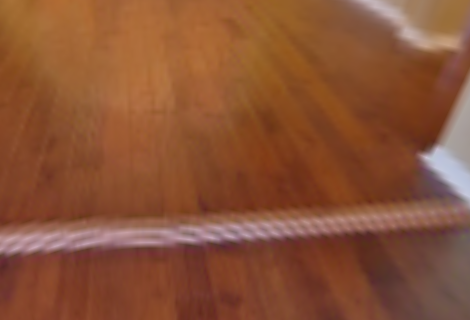}}

\subfloat[(d) No Perceptual]{
\includegraphics[width=0.325\linewidth]{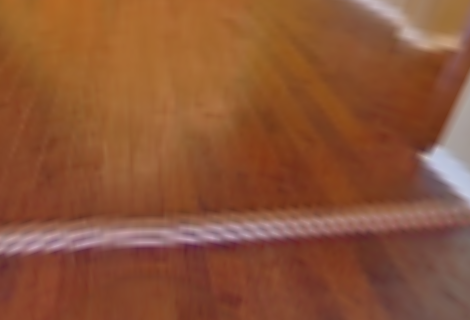}}
\subfloat[(e) No Exclusion]{
\includegraphics[width=0.325\linewidth]{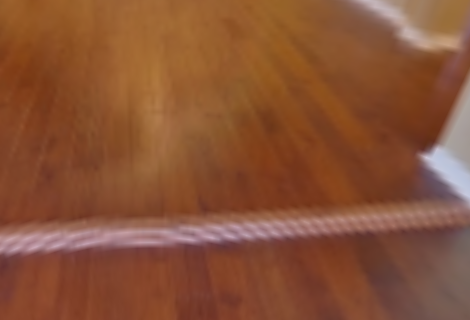}}
\subfloat[(f) Full method]{
\includegraphics[width=0.325\linewidth]{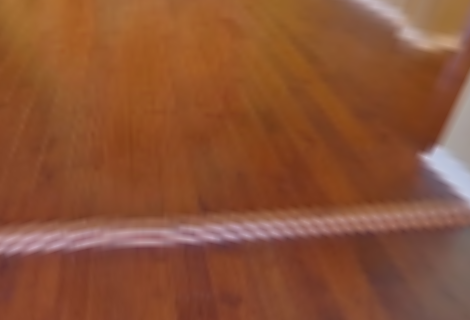}}
    \caption{Ablation study for specular reflection removal by not including adversarial, reconstruction, perceptual or exclusion losses. While quantitative metrics are averaged over the image and over the dataset, the visuals illustrate the subtle but noticeable residual specular reflections when the different losses are removed from our method.}
    \label{fig:loss}
\end{figure}

%% file: fig_compspec.tex
\begin{figure}
    \centering
    \subfloat[(a) Original Image]{
    \includegraphics[width=0.49\linewidth]{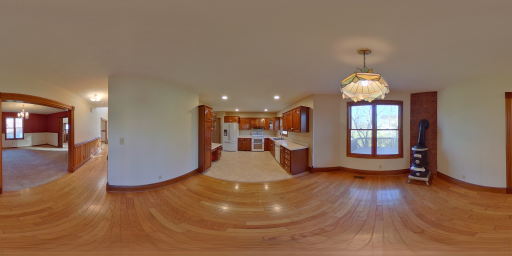}}
    \subfloat[Crop 1]{
    \includegraphics[width=0.241\linewidth]{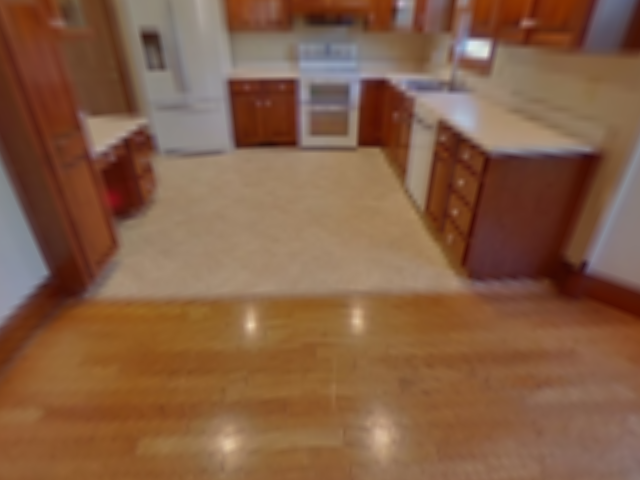}}
    \subfloat[Crop 2]{
    \includegraphics[width=0.241\linewidth]{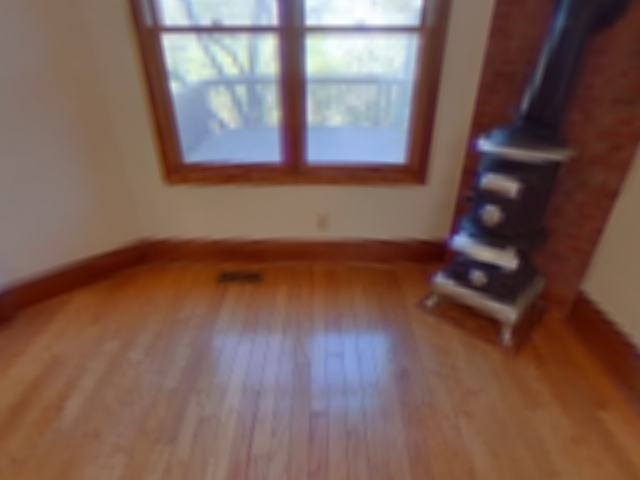}}
    
    \subfloat[(b) LASIRR~\shortcite{dong2021location}]{
    \includegraphics[width=0.241\linewidth]{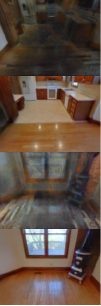}}
    \subfloat[(c) IBCLN~\shortcite{li2020single}]{
    \includegraphics[width=0.241\linewidth]{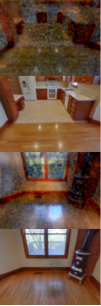}}    \subfloat[(d) JSHDR~\shortcite{dong2021location}]{
    \includegraphics[width=0.241\linewidth]{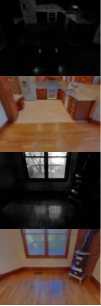}}    \subfloat[(e) Ours]{
    \includegraphics[width=0.241\linewidth]{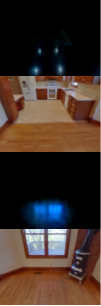}}    \caption{Qualitative results of specular reflection detection and removal. Specular reflections are brightened for visualization. Note that although the walls are shown, only the floor region is evaluated quantitatively. Our method successfully detects and removes the both window and lamp reflections. (d) removes lamp reflections but (b) and (c) do not. (b) (c) (d) leaves residues of window reflections on the floor.}
    \label{fig:compspec}
\end{figure}

%% file: tab_spec_compare.tex
\begin{table}[]
    \caption{Quantitative analysis for specular reflection detection on real data. mIoU (\%) for specular regions are reported. "Ours" is inferred in panoramic domain, while the others are inferred in perspective domain. All methods are evaluated in both domains (cross domain via cropping or merging). JSHDR~\shortcite{fu2021multi} performs better than the other relevant methods, possibly because of a smaller domain gap.}

    \centering
    \scalebox{0.805}{
    \begin{tabular}{lccc}\toprule
    Method & Perspective mIoU $\uparrow$ & Panoramic mIoU $\uparrow$ \\
    \midrule
    BF~\cite{yang2010real} &11.3 &8.9\\
    IBCLN~\cite{li2020single} & 15.7&15.0\\
    LASIRR~\cite{dong2021location} & 20.5&17.5\\
    JSHDR~\cite{fu2021multi} & 21.4&14.1 \\
    Ours  &  \textbf{41.8}&\textbf{38.9}\\
    \bottomrule
    \end{tabular}}

    \label{tab:spec_compare}
\end{table}

%% file: tab_diff_compare.tex
\begin{table}[]
    \centering
    \caption{Quantitative analysis for specular reflection removal on synthetic data. "Ours" and "IBCLN (re-train)" are inferred in panoramic domain, while the other methods are inferred in perspective domain. All methods are evaluated in both domains (cross domain via cropping or merging). To bridge the domain gap, we re-train IBCLN~\shortcite{li2020single} on our data. Although performance is improved, it does not reach the level of our method.}
    \scalebox{0.805}{
    \begin{tabular}{llccccc}\toprule
    \multirow{2}*{\tabincell{c}{Evaluation \\ Domain}}&
   \multirow{2}*{Method} & \multicolumn{2}{c}{All Testdata} & \multicolumn{2}{c}{Strong Reflection}  \\ \cmidrule{3-6}&& PSNR $\uparrow$  & LPIPS $\downarrow$ &PSNR$ \uparrow$  &LPIPS $\downarrow$ \\
    \midrule
    Perspective & BF~\shortcite{yang2010real} & 23.4  & 0.1632 & 17.7 & 0.2326\\
     & IBCLN~\shortcite{li2020single} & 27.5 & 0.1013 & 23.7 & 0.1460\\
     & LASIRR~\shortcite{dong2021location} & 20.8 & 0.1984 & 17.5 & 0.2515\\
    &  JSHDR~\shortcite{fu2021multi} & 32.0 & 0.0354 & 24.1 & 0.1158\\
     & IBCLN~\shortcite{li2020single} (re-train) & 33.4 & 0.0279 & 25.3 & 0.1016\\
    &  Ours &\textbf{34.5} & \textbf{0.0271} & \textbf{27.0} & \textbf{0.0935} \\
    \midrule
    Panoramic & BF~\shortcite{yang2010real} & 22.1  & 0.1526 & 16.5 & 0.1934\\
     & IBCLN~\shortcite{li2020single} & 28.8 & 0.0890 & 24.2 & 0.1418\\
    &  LASIRR~\shortcite{dong2021location} & 23.0 & 0.1425 &20.1 & 0.1788\\
    &  JSHDR~\shortcite{fu2021multi} & 30.6 & 0.0698 & 22.5 & 0.1499\\
    &  IBCLN~\shortcite{li2020single} (re-train) & 32.3 & 0.0250 & 24.7 & 0.0743\\
    &  Ours &\textbf{33.3} & \textbf{0.0241} & \textbf{25.3} & \textbf{0.0676} \\
    \bottomrule
    \end{tabular}}

    \label{tab:diff_compare}
\end{table}

%% file: fig_highres.tex
\begin{figure*}
    \centering
    \subfloat[(a) Original Image]{
    \includegraphics[height=0.6\linewidth]{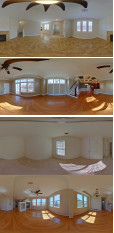}}
    \subfloat[(b) Lighting Effects (Brightened)]{
    \includegraphics[height=0.6\linewidth]{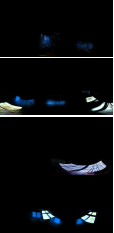}}
    \subfloat[(c) Effects Removed]{
    \includegraphics[height=0.6\linewidth]{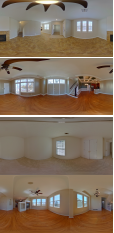}}
    \subfloat[]{
    \includegraphics[height=0.6\linewidth]{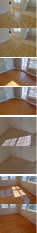}}
    
    \caption{High resolution appearance decomposition results. The lighting effects are successfully detected on a variety of floor types. For specular reflection removal, the floor texture in (c) is consistent with that in (a). Sunlight regions are inpainted with correct colors. Some floor textures are imperfect due to the limitation of the inpainting algorithm, e.g, the third row, where the camera tripod is inpainted in the original image. }
    \label{fig:highres}
\end{figure*}

%% file: fig_special.tex
\begin{figure}
    \centering
    \subfloat[(a) RGB]{
    \includegraphics[width=0.49\linewidth]{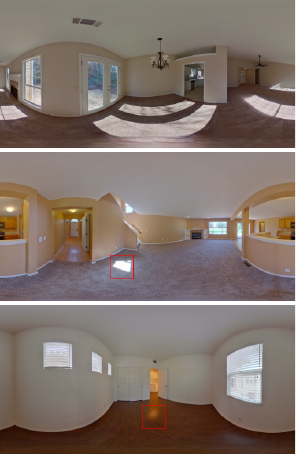}}
    \subfloat[(b) Lighting Effects (brightened)]{
    \includegraphics[width=0.49\linewidth]{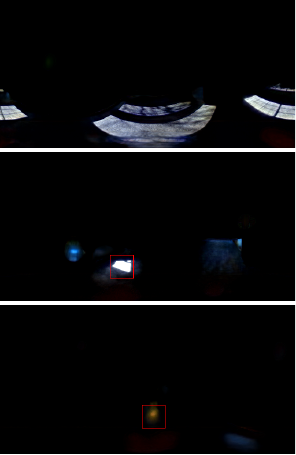}}
    \caption{Diverse lighting conditions of testing images. Row 1: large sunlight region; Rows 2-3: effects from (partially) occluded light sources (see red boxes in (a)). Our method successfully detects the lighting effects.}
    \label{fig:special}
\end{figure}

%% file: sec8_apps.tex
\section{Applications}
\label{sec:apps}
\input{fig_albedo}

\paragraph{Improvement of Albedo Estimation.}
Albedo estimation via intrinsic image decomposition suffers from the existence of specular reflection and direct sunlight. As shown in Fig.~\ref{fig:albedo} (b), a pre-trained USI3D~\cite{liu2020unsupervised} incorrectly bakes specular reflection and direct sunlight into the albedo component. When we run the same algorithm on the ambient image after removing the lighting effects, the albedo estimation (c) is signifiantly better.

\paragraph{Improvement of Sun Direction Estimation.}
\label{ssec:sundir}
With the detected direct sunlight, we can improve the sun direction estimation proposed in Sec.~\ref{ssec:coarsemask}. There are two modifications: (1) Instead of using the RGB image, we use the direct sunlight image for matching; (2) The wall sunlight is also considered. Previously we do not consider walls for coarse sun direction estimation because walls are usually white, which may mislead the matching score. 

\input{fig_sundir}

To evaluate the sun direction, we obtained sun elevation angles for part of the dataset from the authors of ZInD~\cite{cruz2021zillow}. The elevation angles are calculated based on timestamps and geolocations. The ground truth azimuth angles are not available because of unknown camera orientations. The elevation angles provided may not be always accurate because the timestamps are based on the time the data were uploaded to the cloud. Thus, we manually filter obviously incorrect values and select images with visible sunlight for evaluation. A total of 257 images are evaluated.

\input{fig_changemat}
\input{fig_changedir}
\input{fig_apps}

We search for the top-1 direction at step size 1$^\circ$. Fig.~\ref{fig:sundir} plots the histogram of the angular estimation error. More than half of the estimations are within 10$^\circ$ error. Fig.~\ref{fig:sundir} also visualizes an example of projecting the window mask according to the estimated sun direction. Compared with the coarse sun direction, the improved fine estimation provides more accurate estimations.

\paragraph{Changing Floor Material.}
The decomposition is used to change flooring with texture and BRDF parameters. Examples include switching between wood (specular), carpet (diffuse) or tile (specular). Fig.~\ref{fig:changemat} shows four examples, including wood-to-wood (keep specularity), wood-to-carpet (remove specularity), carpet-to-wood (render specularity via HDR map), and carpet-to-carpet changes. Direct sunlight are rendered on the new material by scaling the floor intensity according to the estimated sunlight scale.

\footnotetext[7]{Texture credits: Column 1 wood\copyright nevodka.com/Adobe Stock, carpet\copyright anya babii/Adobe Stock; Column 2 wood\copyright Kostiantyn/Adobe Stock, carpet\copyright Paul Maguire/Adobe Stock.}

\paragraph{Changing Sun Direction.}
Using the estimated sun direction, we project the sunlit region back to the window position. Then, we re-project sunlight onto the floor and the walls using a new sun direction. Fig.~\ref{fig:changedir} visualizes two examples of changing sun direction.

\paragraph{Furniture Insertion.}
With the decomposition result, sun direction and the estimated HDR map (Sec.~\ref{ssec:implement}), we can insert virtual objects into an empty room, by calculating the occluded specular reflection and direct sunlight via ray tracing. Using the computed layout (scale set manually),  we render ambient, sunlight and specular effects separately and combine them. Fig.~\ref{fig:apps} shows several insertion examples, where sunlight is cast on the desk (a) and the bed (b\&c), specular reflection is blocked by the chair (a) and bed-side table (c), and sunlight is blocked by the clothing hanger (b) and bed (c).

\input{fig_compinsert}

Fig.~\ref{fig:compinsert} shows scenes rendered with Li~\etal~\cite{li2020inverse}, rendered with the estimated HDR map without appearance decomposition, and rendered with our full method. Li~\etal~\cite{li2020inverse} trains on synthetic data with ground truth supervision and predicts illumination and scene properties for a perspective view. The method estimates spatially-varying lighting for each pixel rather than each 3D voxel. A selected pixel and its lighting is used for rendering. This fails when the object parts are far away from the selected pixel in 3D (\eg, first row in Fig.~\ref{fig:compinsert}). Besides, the method renders shadows in the specular region by scaling the original floor intensity without explicitly removing blocked specular reflection, leading to insufficient darkening of the floor (\eg, second row).

\paragraph{Combination of Multiple Applications.}
The applications can be combined to allow more flexible virtual staging effects. Fig. ~\ref{fig:apps} shows high-quality visualizations that include two or more of changing floor material, sun direction and object insertion.

\footnotetext[8]{Credits: (a) landscape painting\copyright adobestock3d/Adobe Stock,
ship painting\copyright Rijksmuseum/Adobe Stock,
desk\copyright Francesco Milanese/Adobe Stock,
computer\copyright Kamil/Adobe Stock,
shelf\copyright adobestock3d/Adobe Stock,
chair\copyright Kamil/Adobe Stock,
bookcase\copyright Cassie Cerpa/Adobe Stock,
lamp\copyright Marcus/Adobe Stock,
plant\copyright Kids Creative Agency/Adobe Stock;
(b) coat rack\copyright Alexandre/Adobe Stock,
mirror\copyright Francesco Milanese/Adobe Stock, 
dresser\copyright Gregory Allen Brown/Adobe Stock,
notebook\copyright HQ3DMOD/Adobe Stock,
clock\copyright Francesco Milanese/Adobe Stock,
bedside table\copyright Jacob Rose/Adobe Stock,
bed\copyright shabee/Adobe Stock.
(c)-(h): See Fig.~\ref{fig:framework} and Fig.~\ref{fig:changemat}.}

%% file: fig_albedo.tex
\begin{figure}
    \centering
    \subfloat[(a) Original Image]{
    \includegraphics[width=0.49\linewidth]{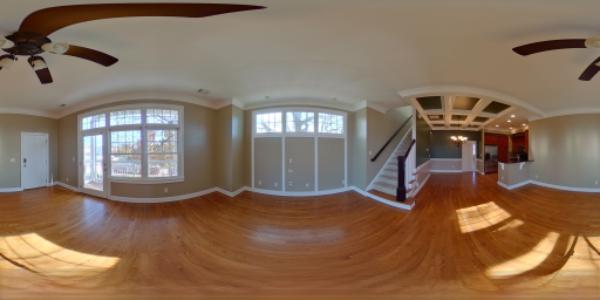}}
    \subfloat[(b) Albedo estimated from (a)]{
    \includegraphics[width=0.49\linewidth]{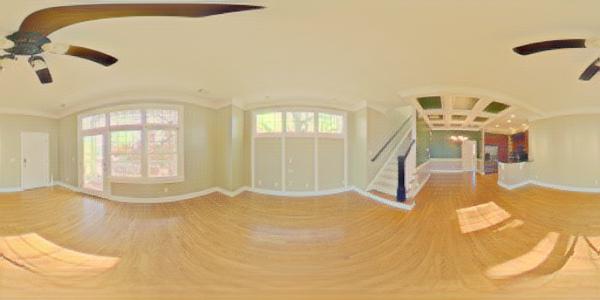}}

    \subfloat[(c) Ambient Image]{
    \includegraphics[width=0.49\linewidth]{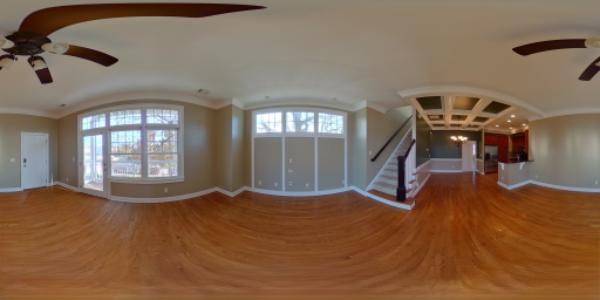}}
    \subfloat[(d) Albedo estimated from (c)]{
    \includegraphics[width=0.49\linewidth]{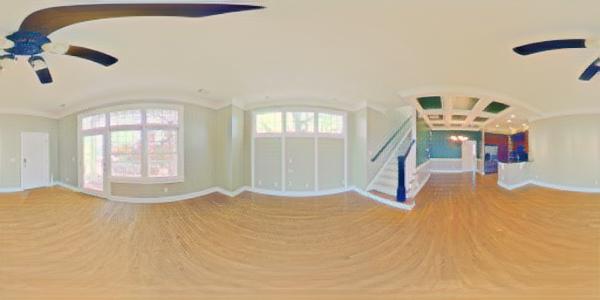}}
    \caption{When an albedo estimation model~\cite{liu2020unsupervised} pre-trained on a mixture of synthetic and real perspective images is applied to our data (a), the domain shift causes obvious errors (b). But when applied to our ambient image (c), the same method is able to estimate albedos more accurately (d), without requiring to retrain on our dataset (which would not be possible in any case given the lack of ground truth annotation for supervision).}
    \label{fig:albedo}
\end{figure}

%% file: fig_sundir.tex
\begin{figure}
    \centering
    \includegraphics[width=0.95\linewidth]{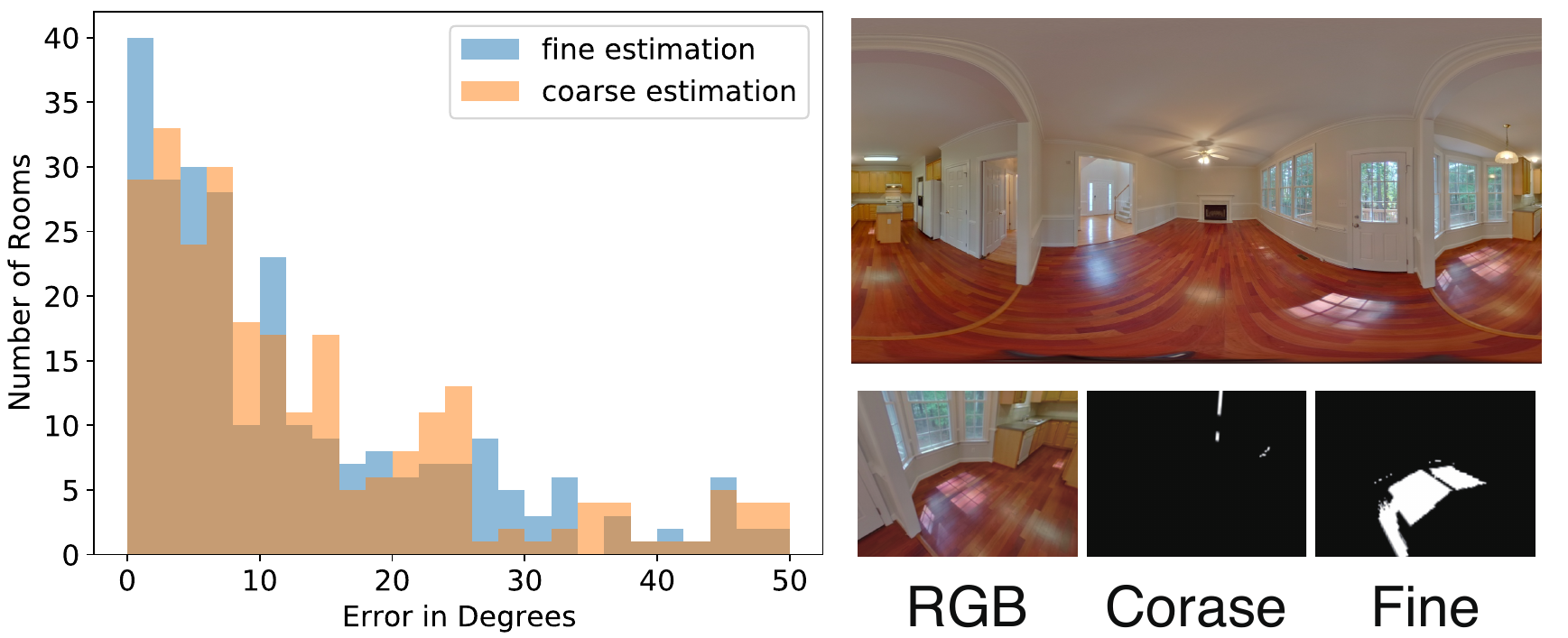}
    \caption{Sun direction estimation results. Left: histogram of sun elevation errors for images with visible sunlight; Right: An example of projecting the window mask to the floor and the walls according to estimated sun direction. The fine sun estimations provide more accurate elevation results. 33\% of the fine estimations and of the 28\% coarse estimations are within $5^\circ$ error. 53\% of the fine estimations and of the 52\% coarse estimations are within $10^\circ$ error. The example shows that the fine estimation produces a reasonable sun direction even with overlapping sunlight and specular reflection. This is an extreme example where coarse estimation completely fails and the fine estimation significantly improves the result.}
    \label{fig:sundir}
\end{figure}

%% file: fig_changemat.tex
\begin{figure}
    \centering
    \includegraphics[width=0.49\linewidth]{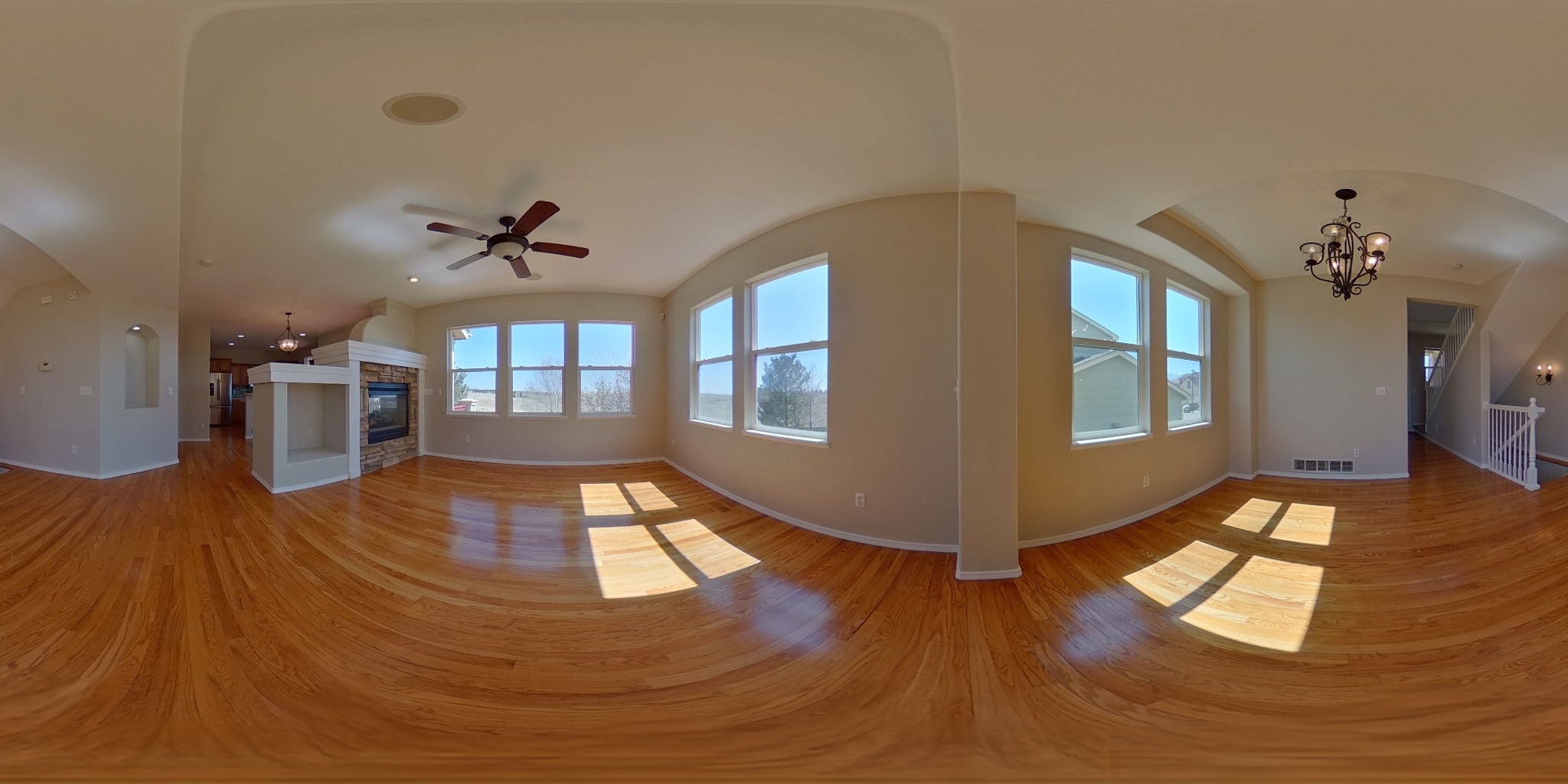}
    \includegraphics[width=0.49\linewidth]{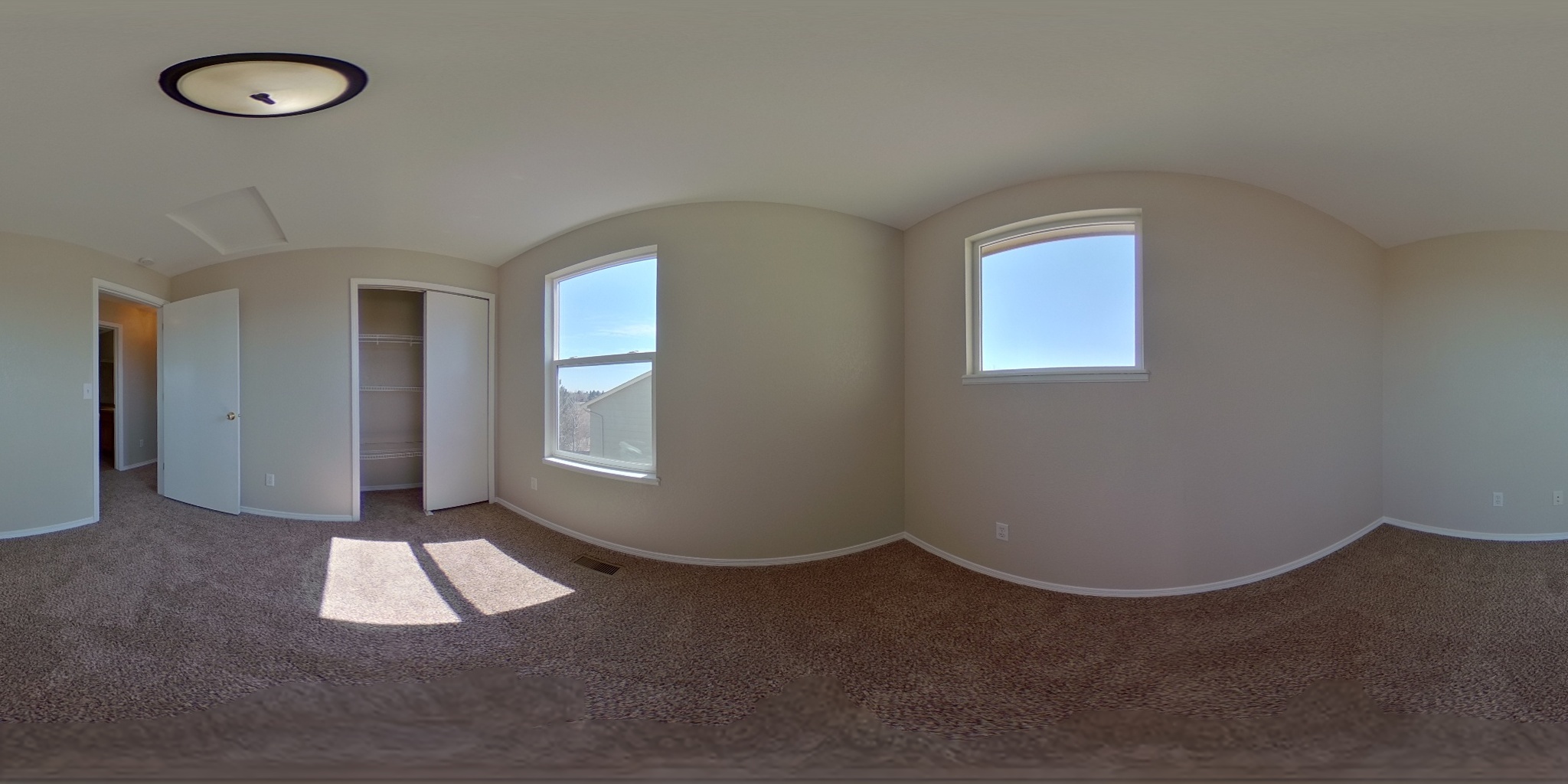}
    
    \includegraphics[width=0.49\linewidth]{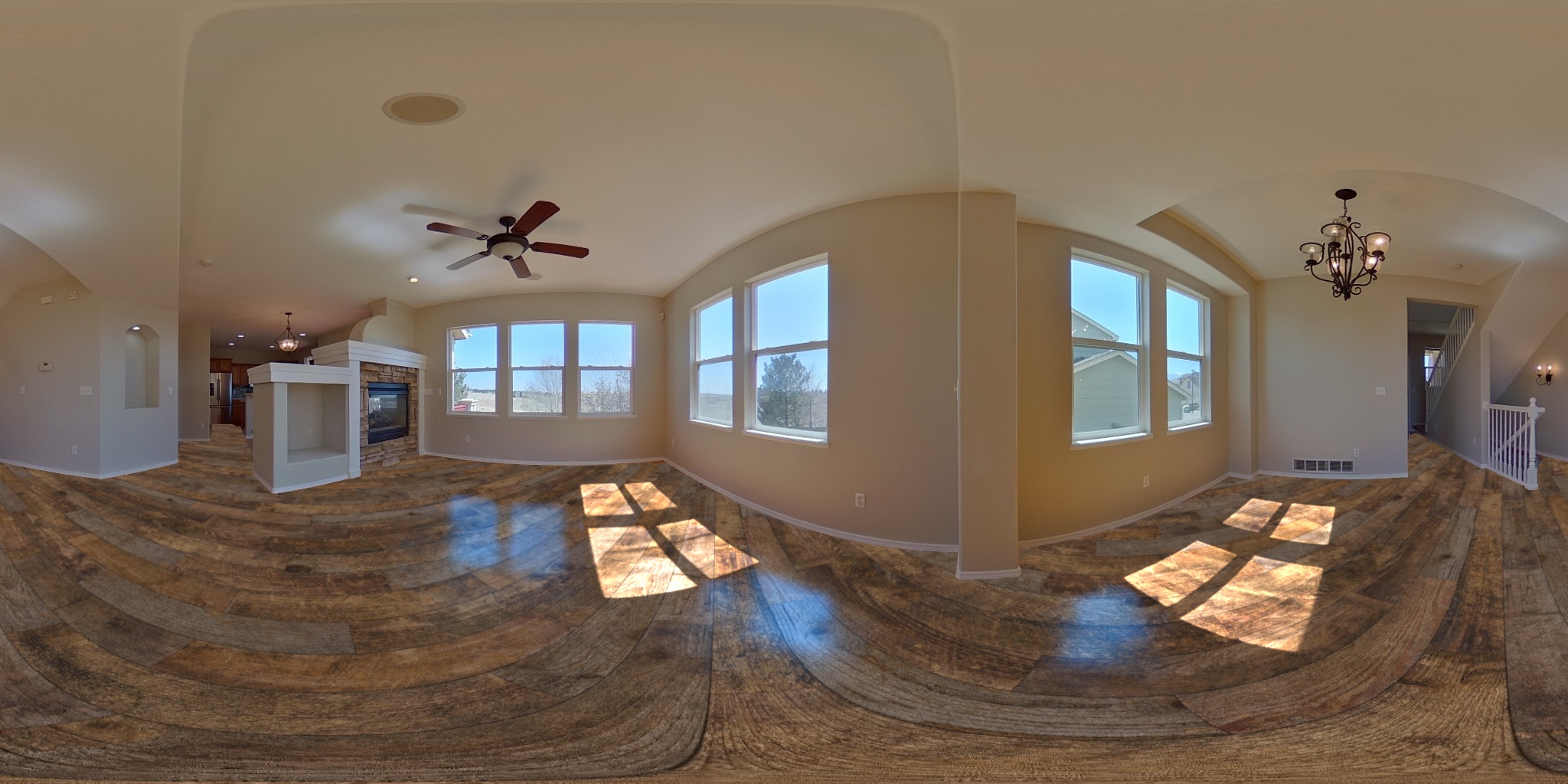}
    \includegraphics[width=0.49\linewidth]{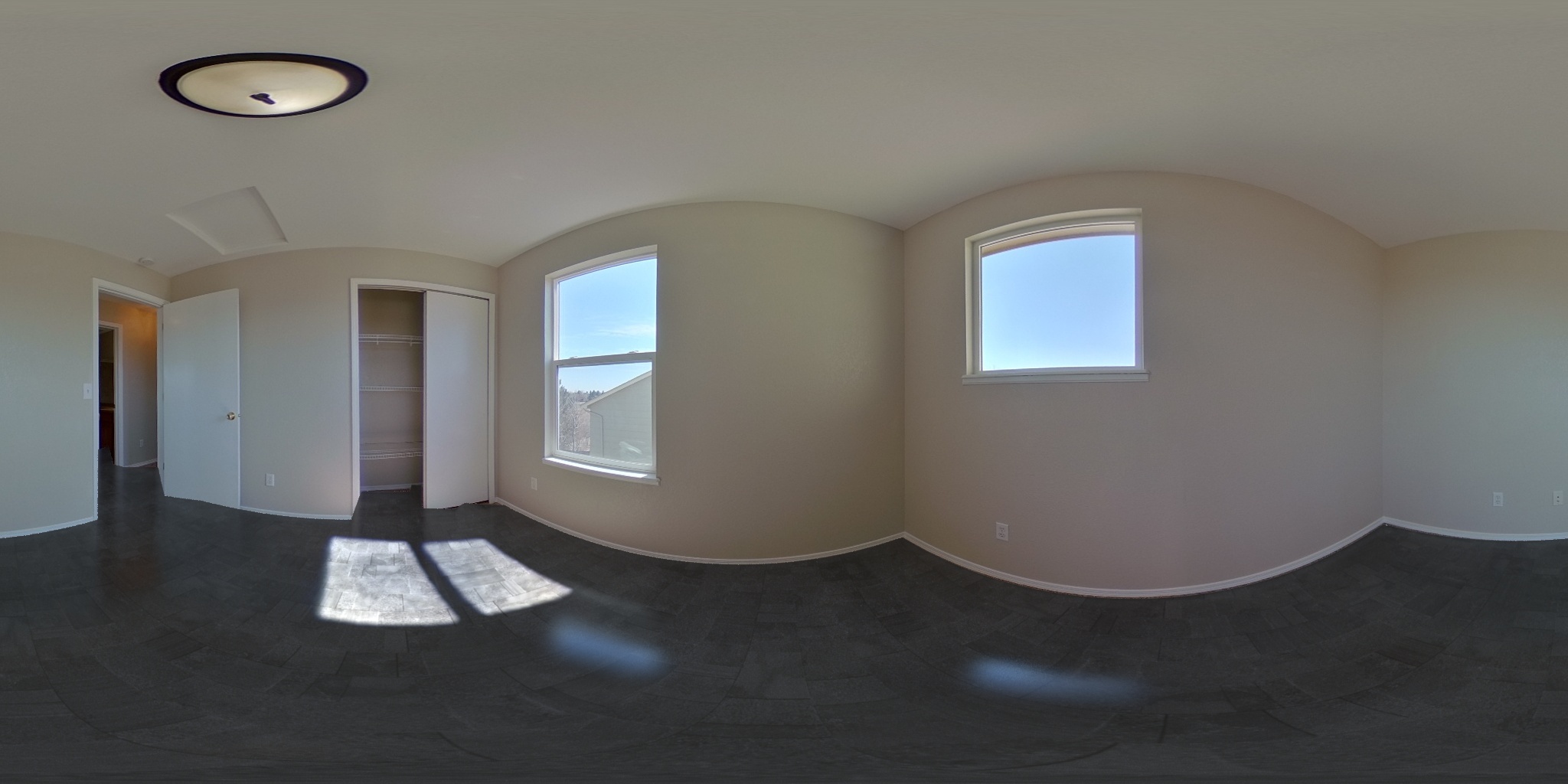}
    
    \includegraphics[width=0.49\linewidth]{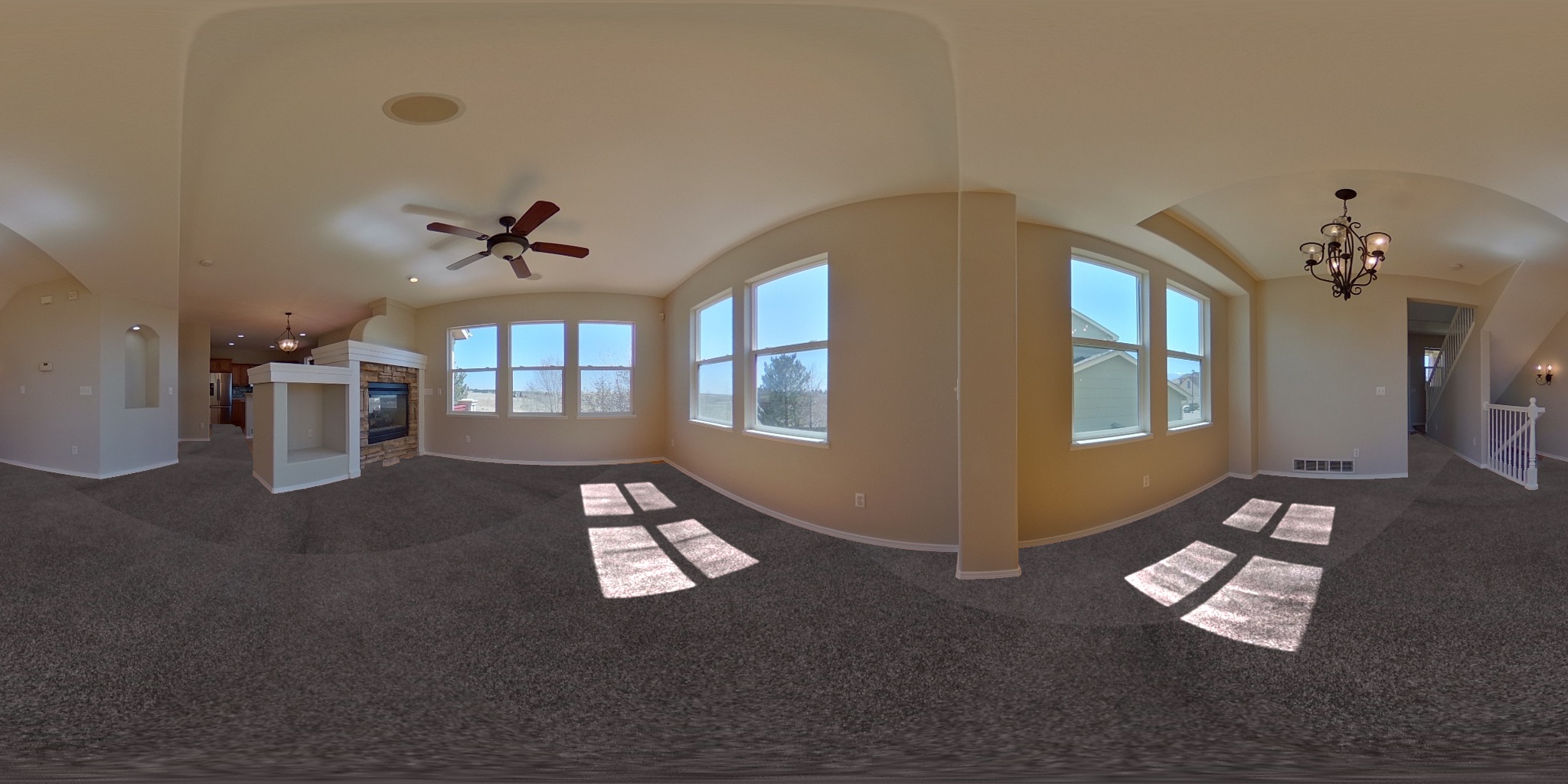}
    \includegraphics[width=0.49\linewidth]{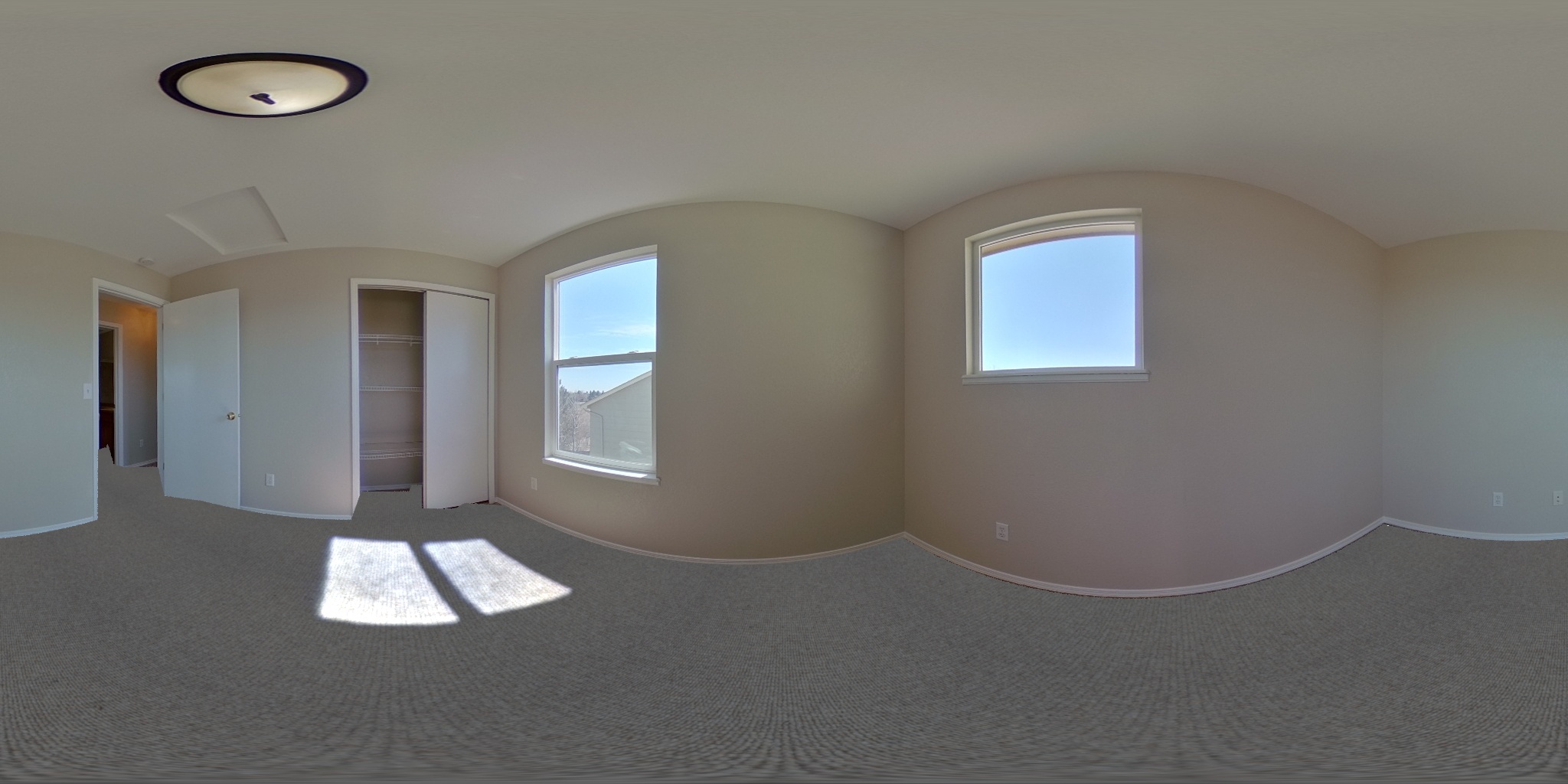}
    \caption{Changing floor material\protect\footnotemark[7]. Row 1: original flooring; Row 2: change to (different) wood; Row 3: change to (different) carpet. The specular reflection on row 2 column 2 is rendered using the estimated HDR map (Sec.~\ref{ssec:implement}) and a uniform surface roughness.} 
    \label{fig:changemat}
\end{figure}

%% file: fig_changedir.tex
\begin{figure}
    \centering
    \includegraphics[width=0.49\linewidth]{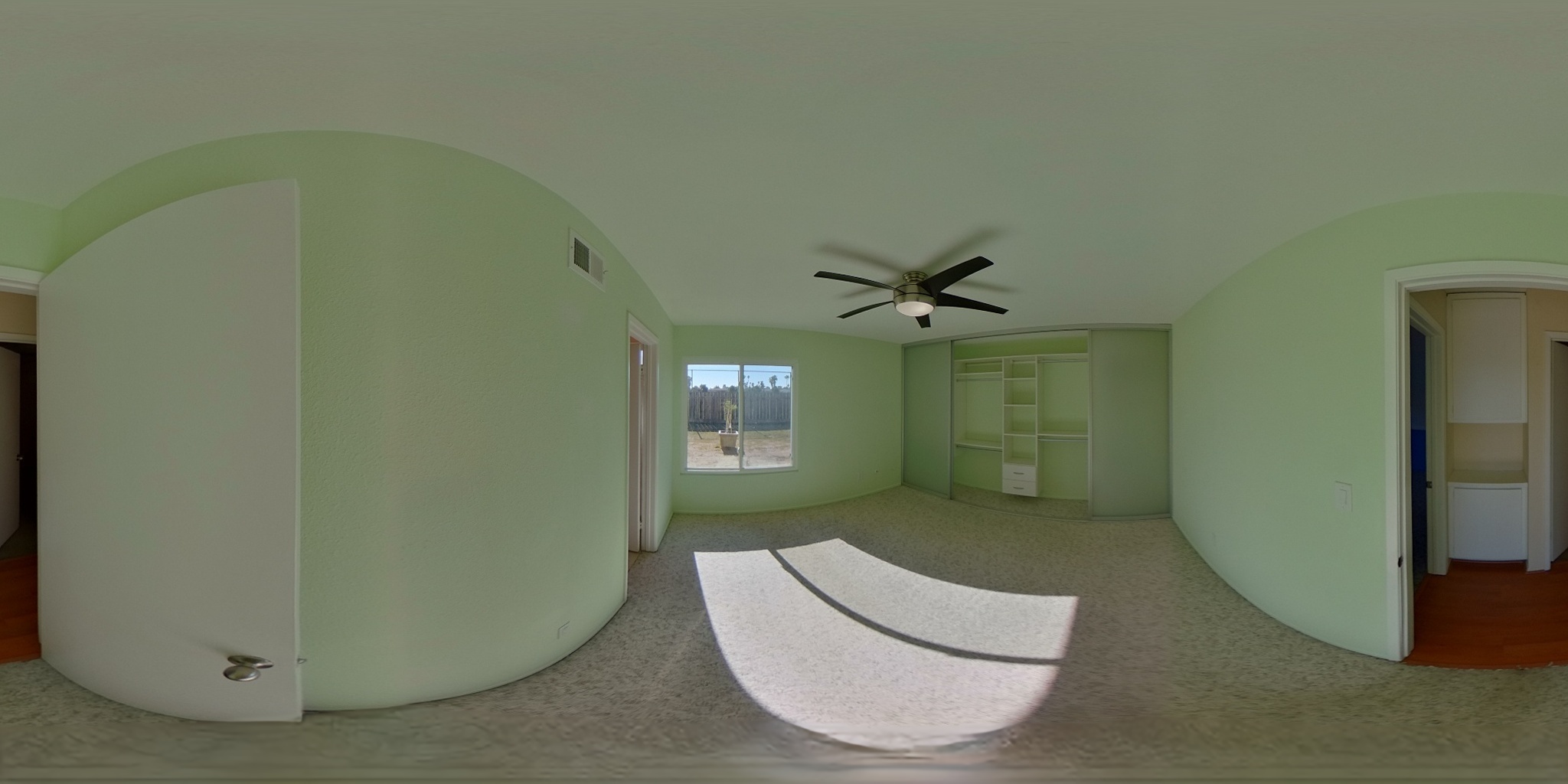}
    \includegraphics[width=0.49\linewidth]{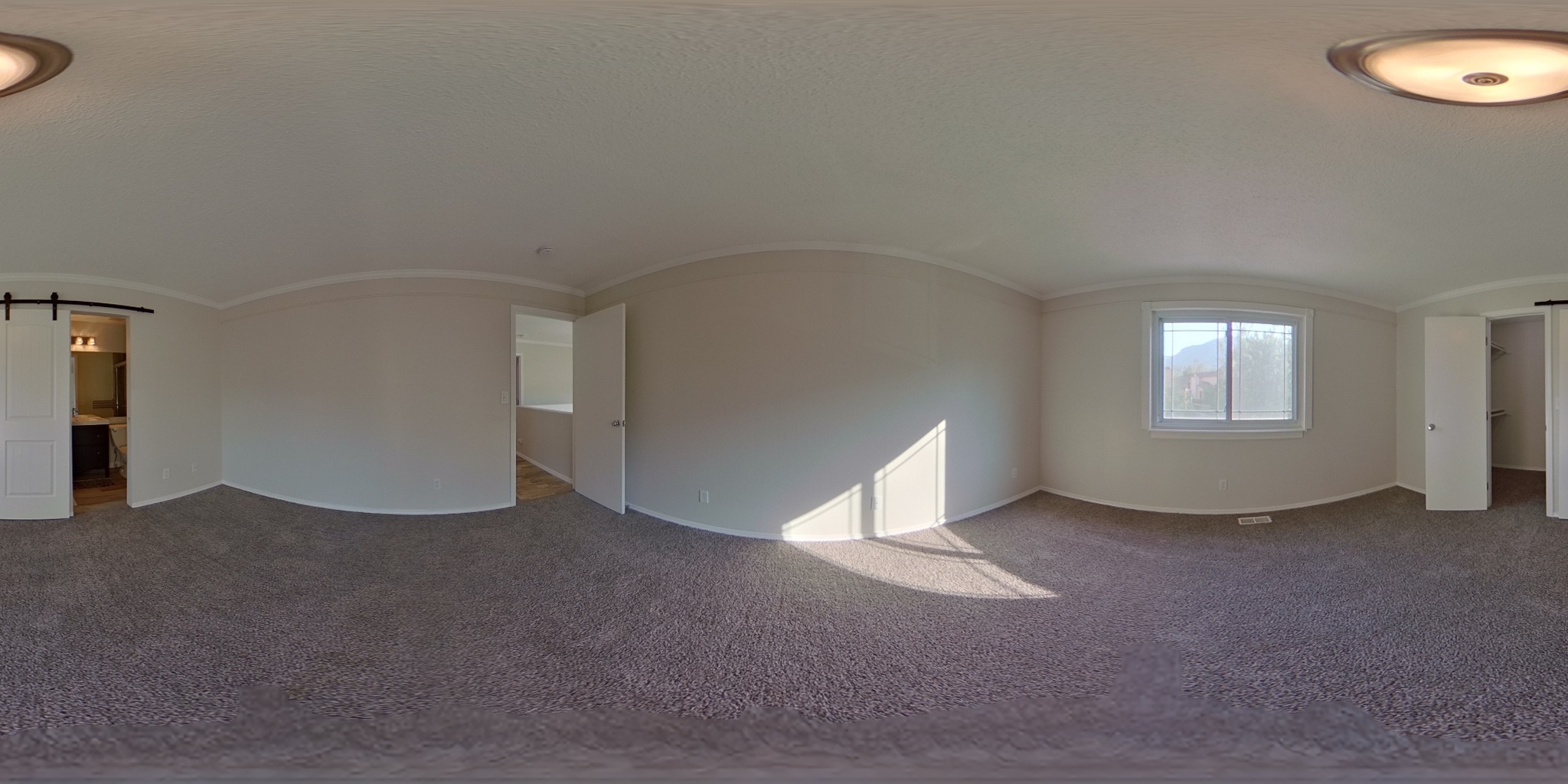}
    
    \includegraphics[width=0.49\linewidth]{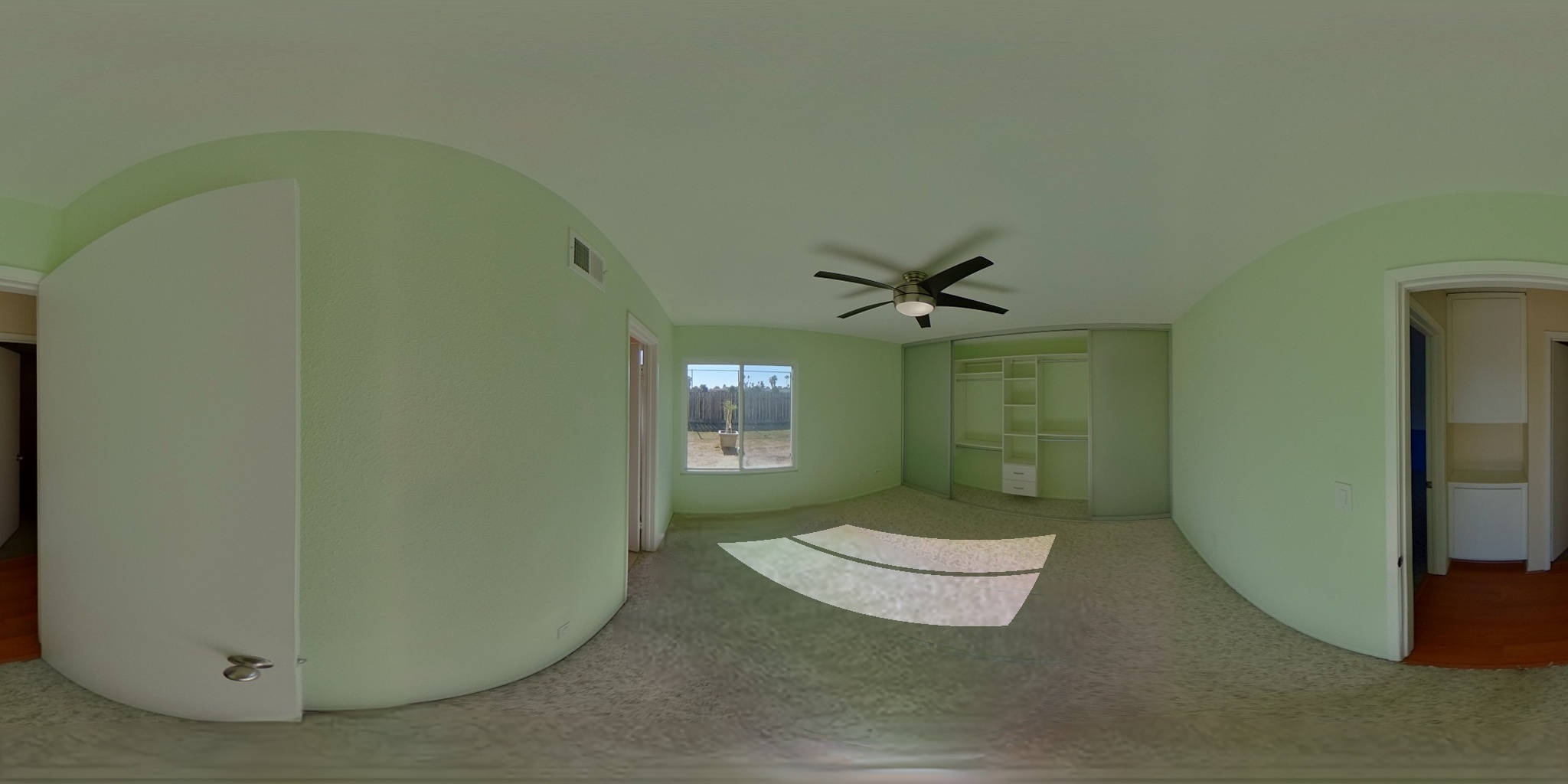}
    \includegraphics[width=0.49\linewidth]{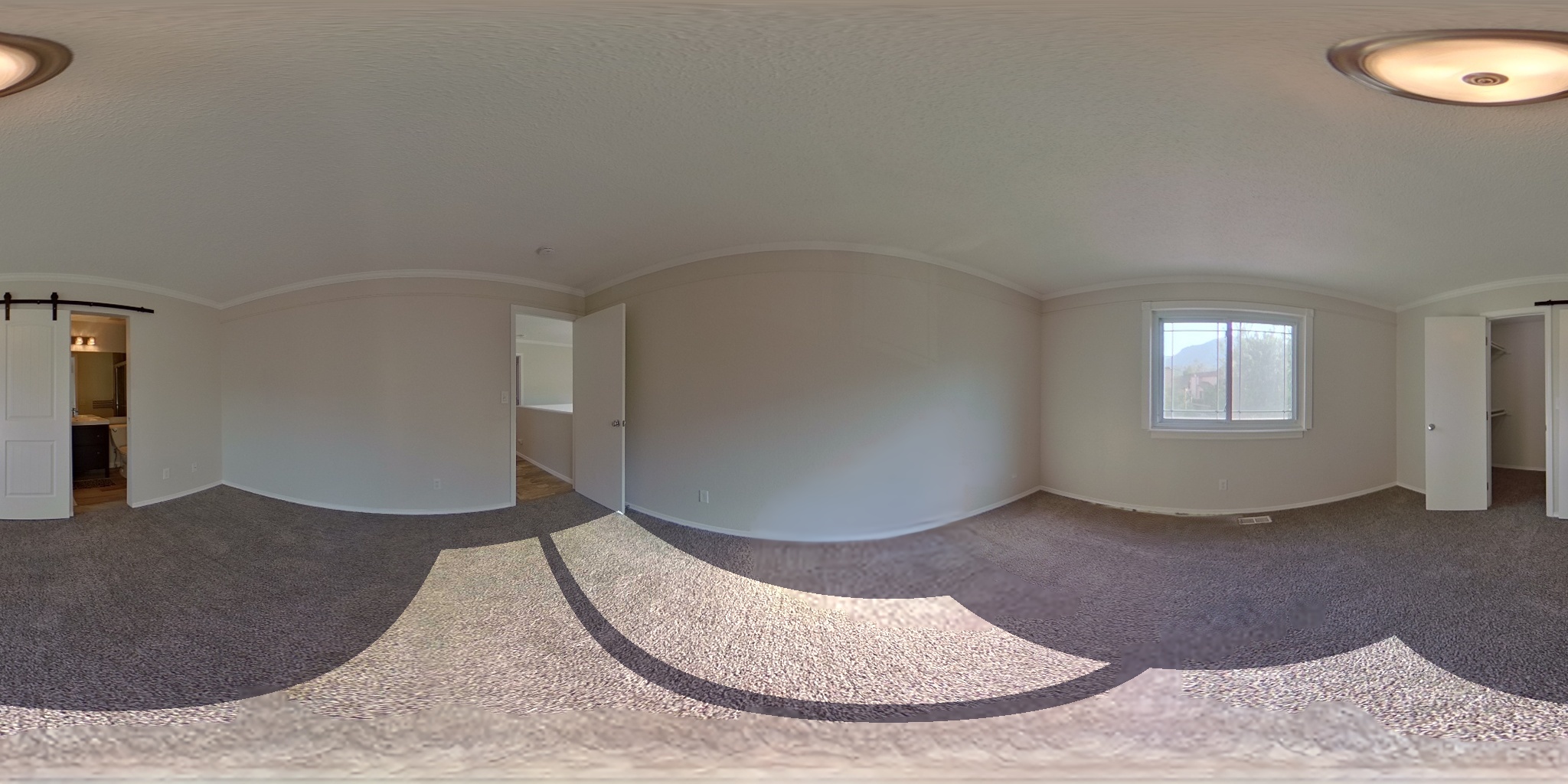}
    
    \includegraphics[width=0.49\linewidth]{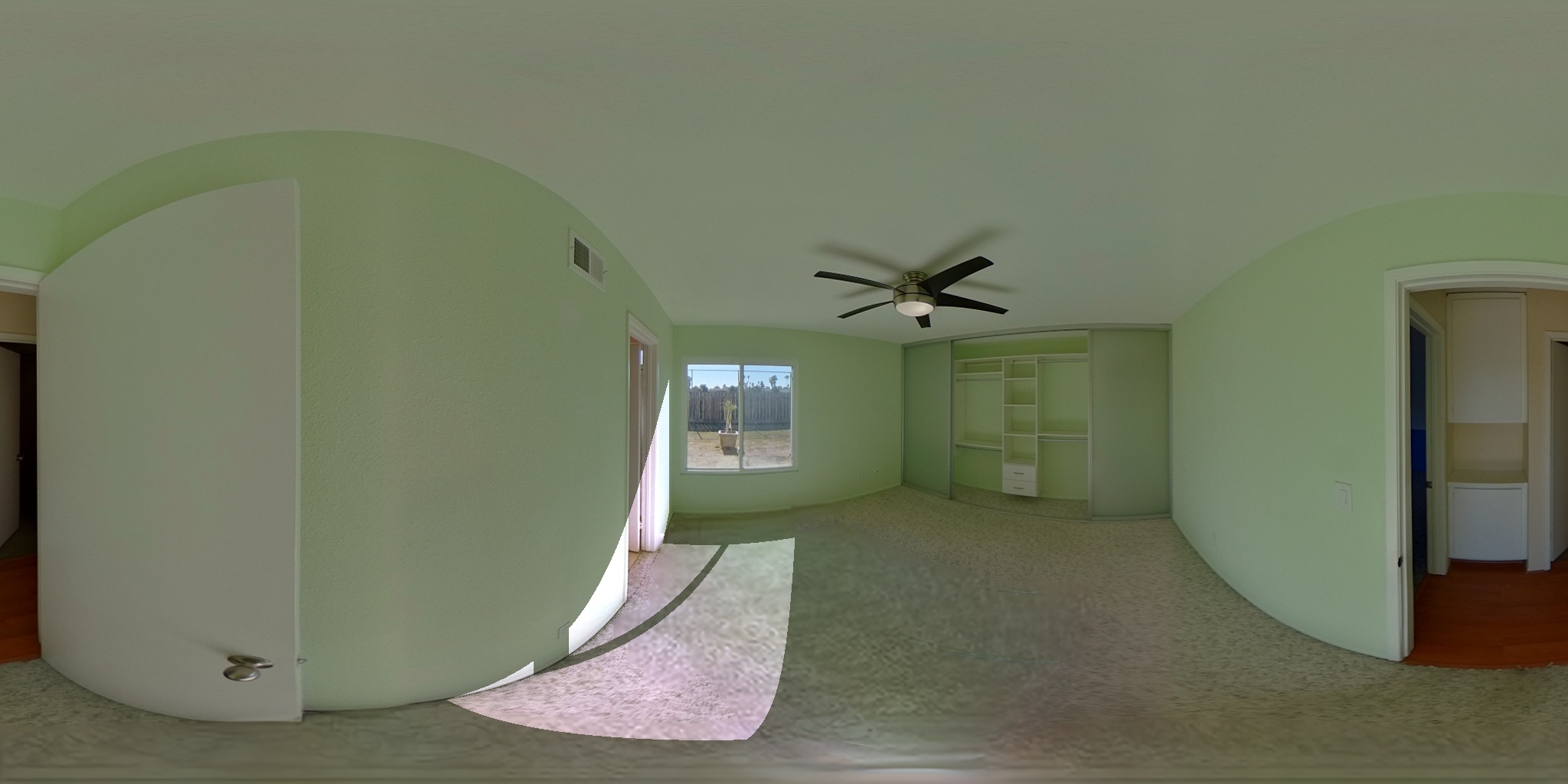}
    \includegraphics[width=0.49\linewidth]{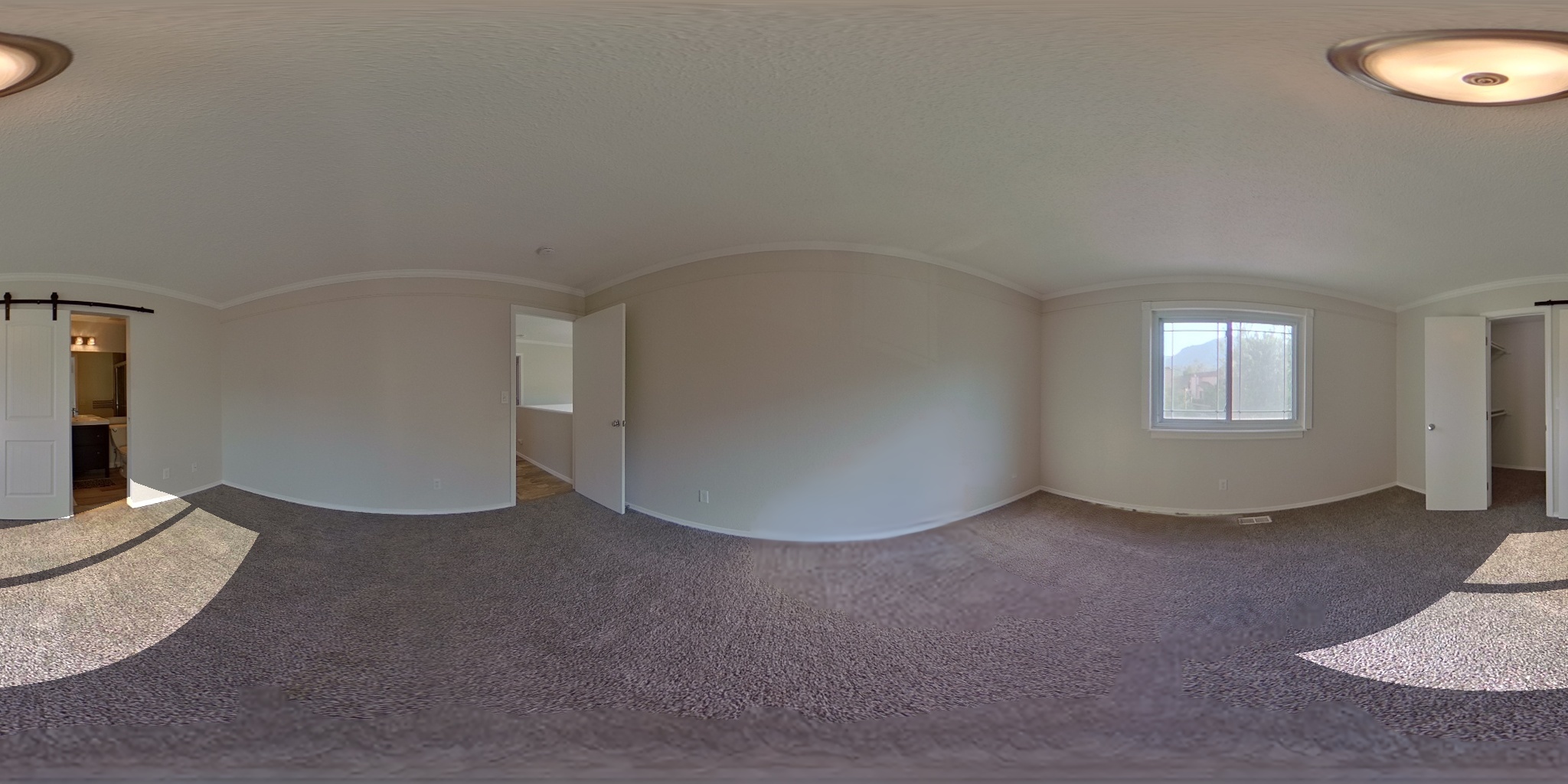}
    
    \includegraphics[width=0.1568\linewidth]{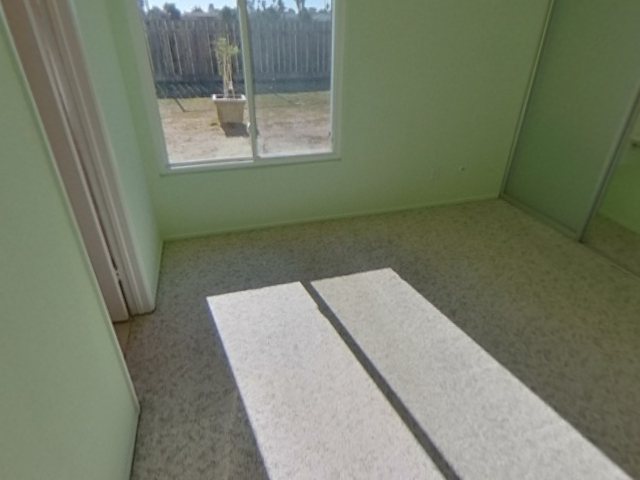}
    \includegraphics[width=0.1568\linewidth]{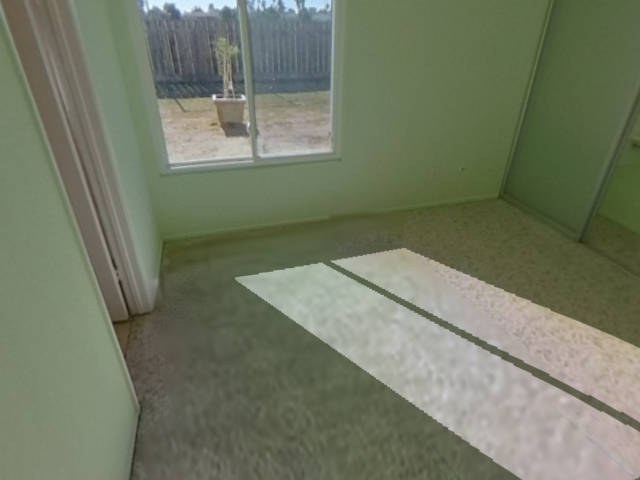}
    \includegraphics[width=0.1568\linewidth]{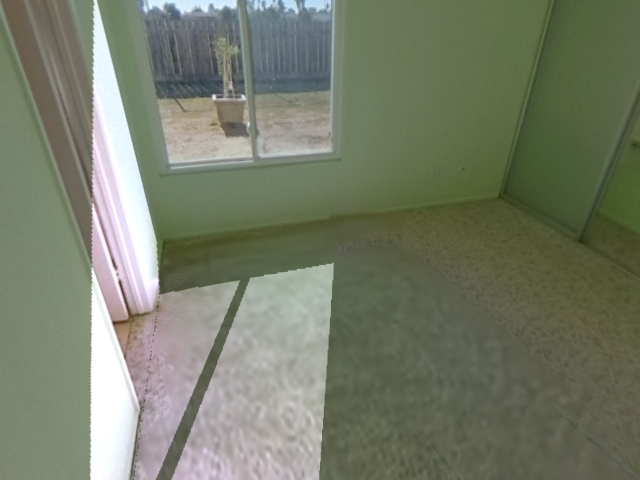}
    \includegraphics[width=0.1568\linewidth]{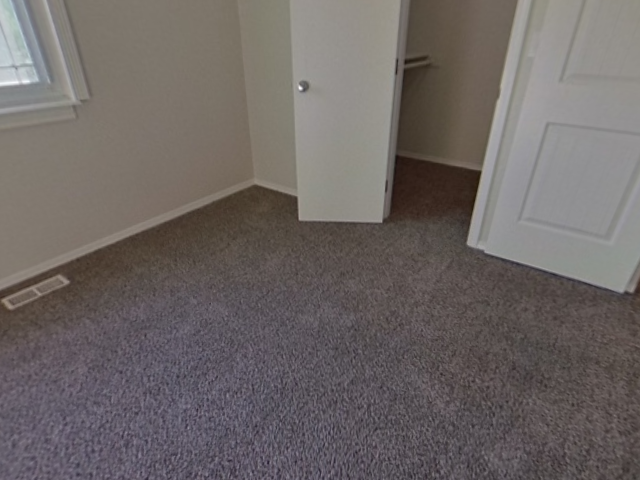}
    \includegraphics[width=0.1568\linewidth]{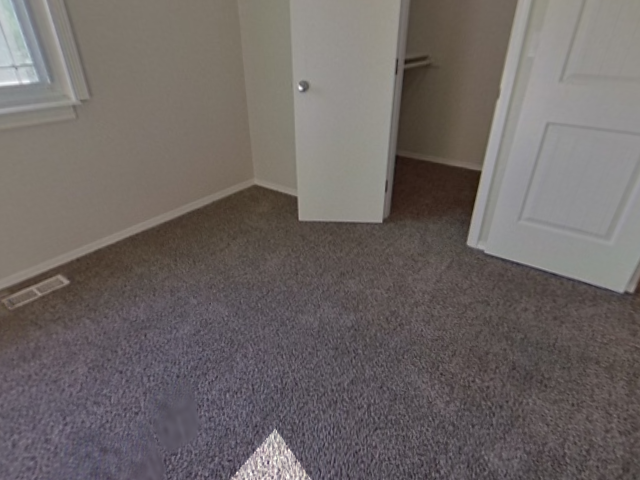}
    \includegraphics[width=0.1568\linewidth]{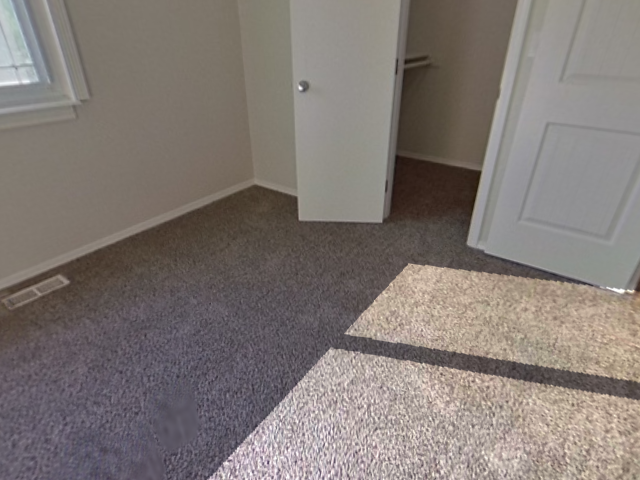}
    
    \caption{Changing sun direction. Row 1: original sun direction; Rows 2-3: New sun directions. Notice the consistency in the shapes and appearances of the sunlit regions in the perspective insets.}
    \label{fig:changedir}
\end{figure}

%% file: fig_apps.tex
\begin{figure*}
\subfloat[(a) Object Insertion for Fig.~\ref{fig:coarsemask}]{
\includegraphics[width=0.49\linewidth]{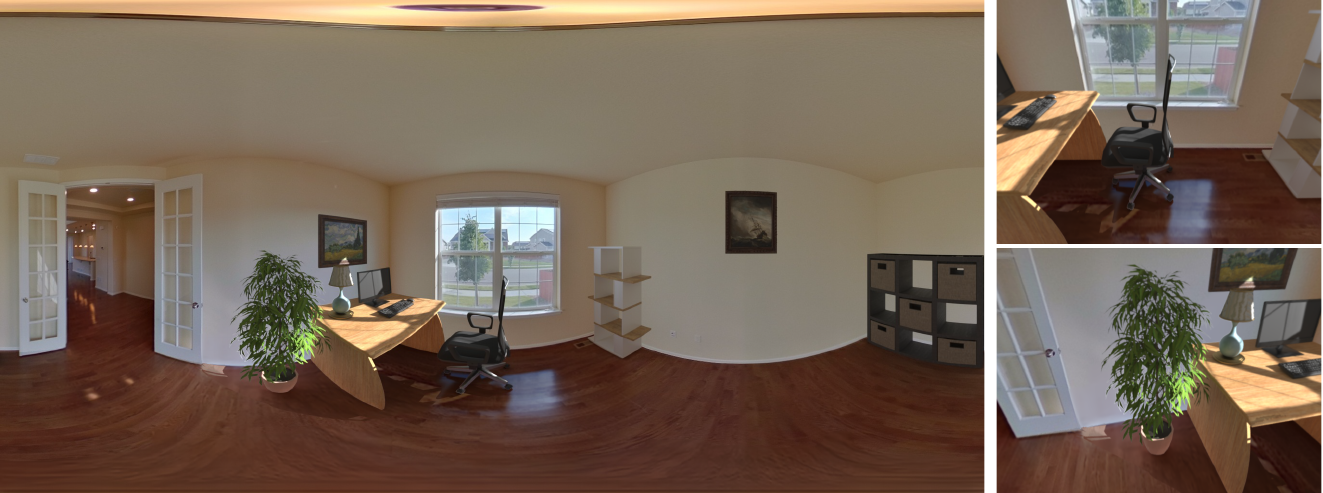}}
\subfloat[(b) Object Insertion for Fig.~\ref{fig:highres} Row 3]{
\includegraphics[width=0.49\linewidth]{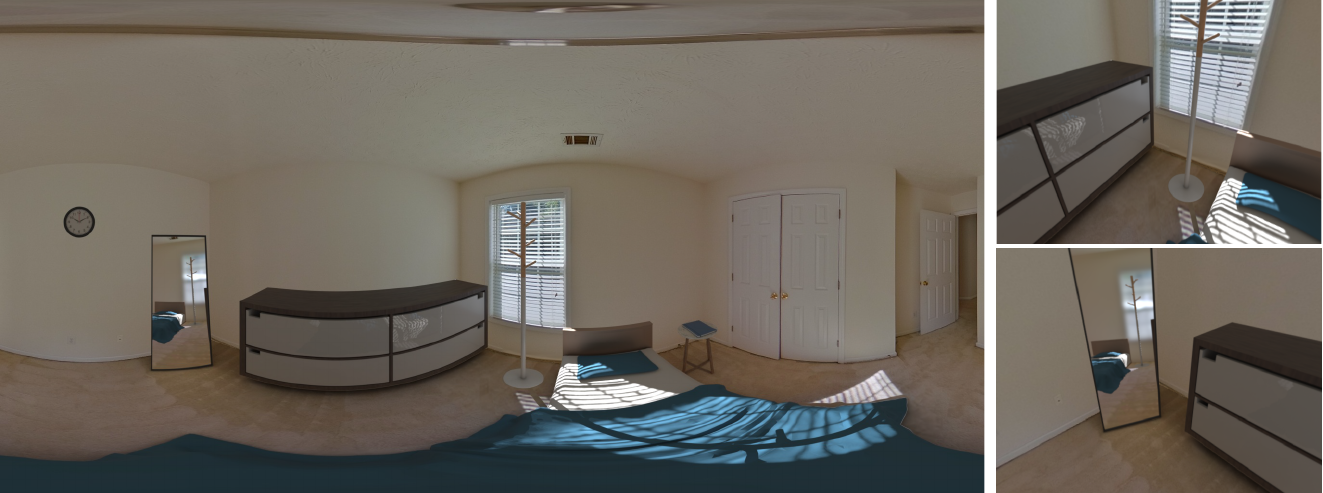}}

\subfloat[(c) Object Insertion for Fig.~\ref{fig:highres} Row 4]{
\includegraphics[width=0.49\linewidth]{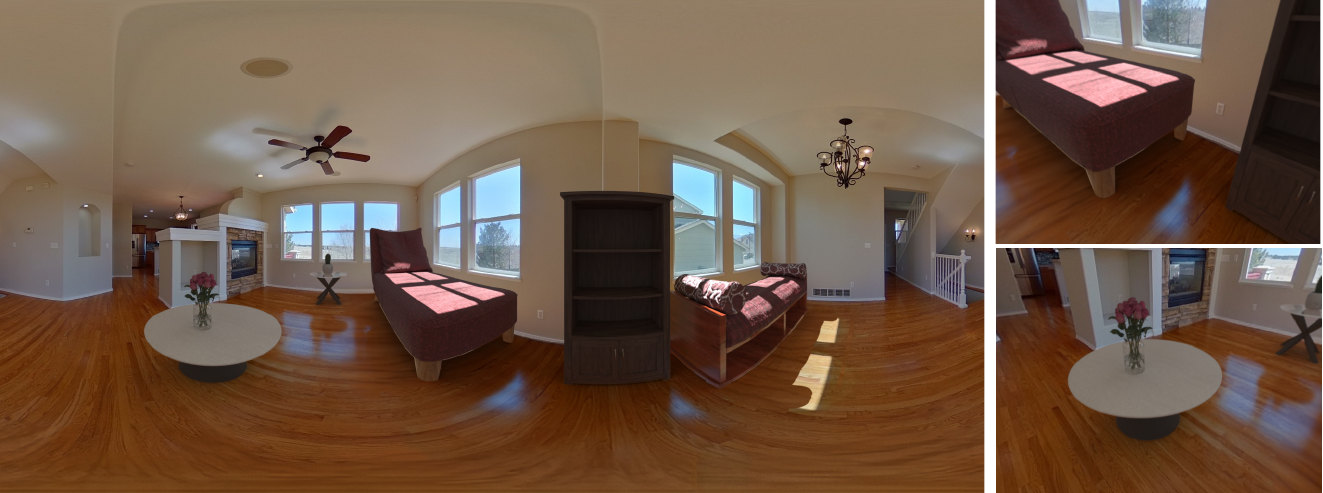}}
\subfloat[(d) Change to Wood + Insertion]{
\includegraphics[width=0.49\linewidth]{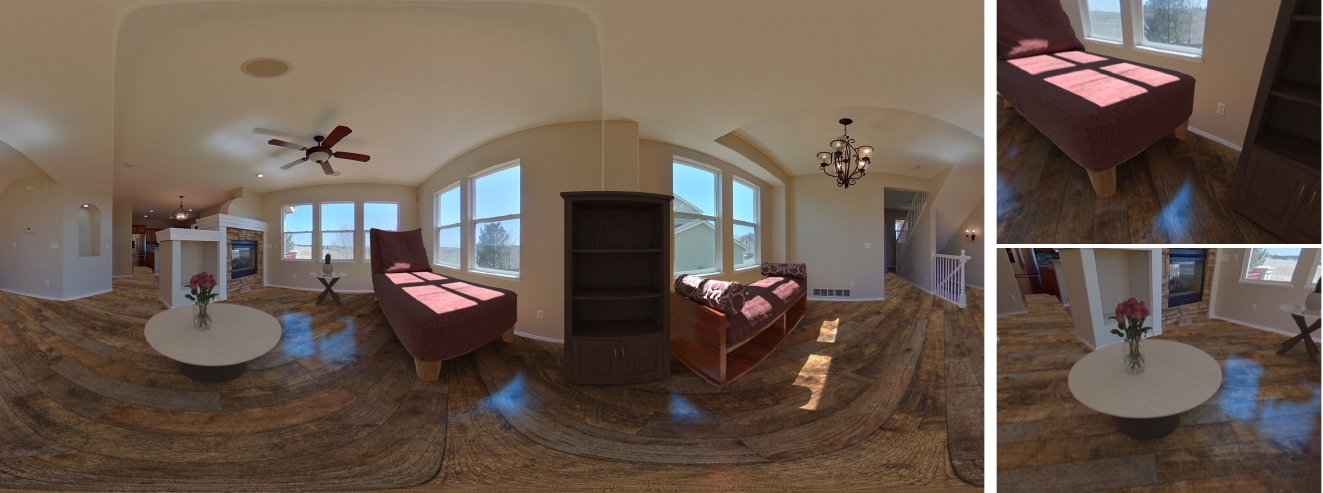}}

\subfloat[(e) Change to Wood + Change Sun Direction]{
\includegraphics[width=0.49\linewidth]{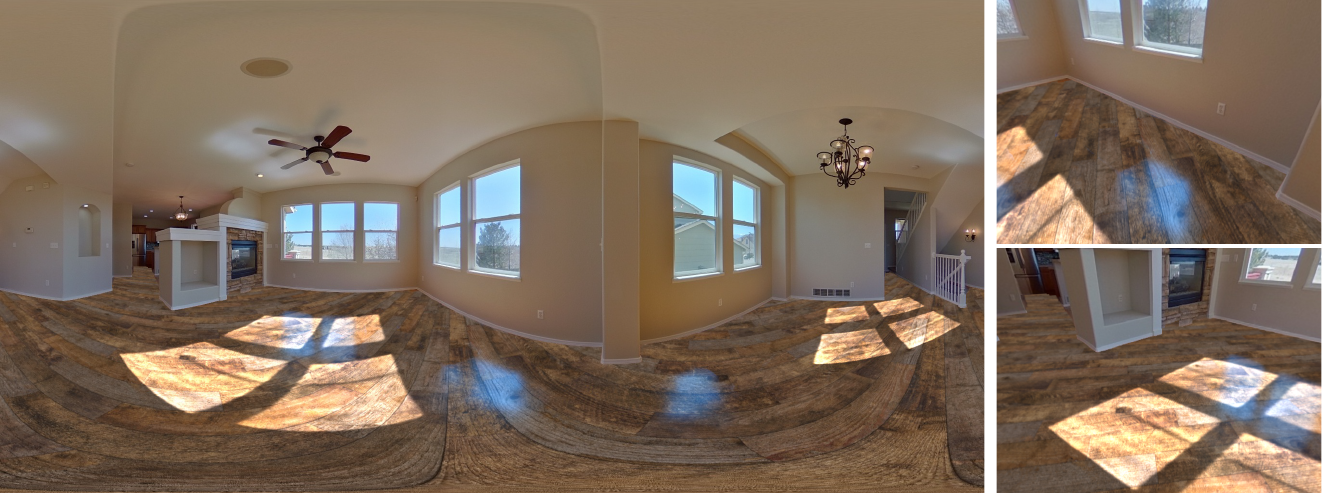}}
\subfloat[(f) Change to Wood + Change Sun Direction + Insertion]{
\includegraphics[width=0.49\linewidth]{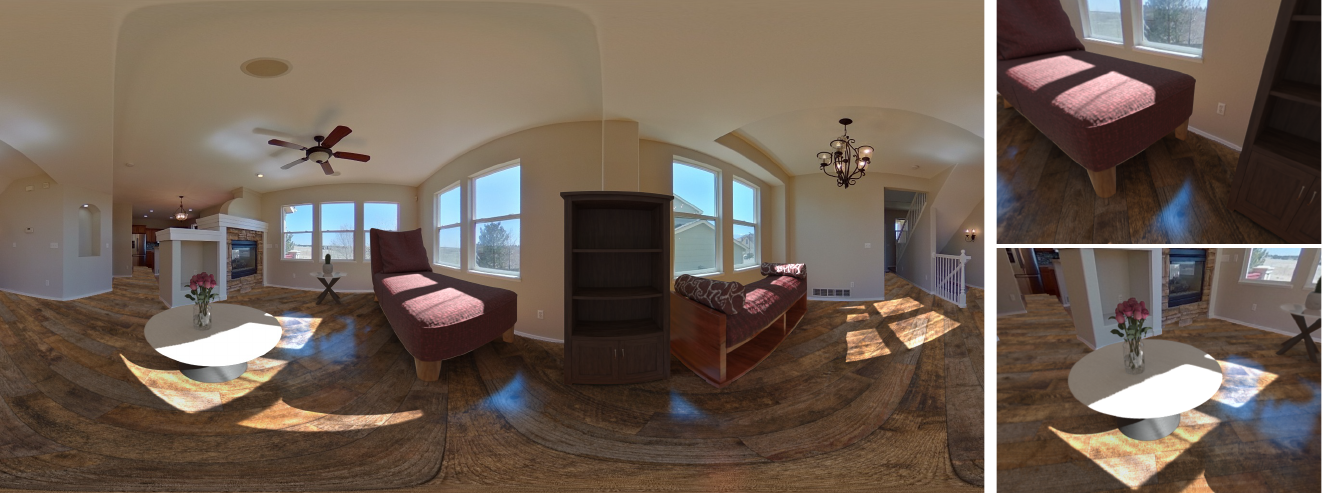}}

\subfloat[(g) Change to Carpet + Insertion]{
\includegraphics[width=0.49\linewidth]{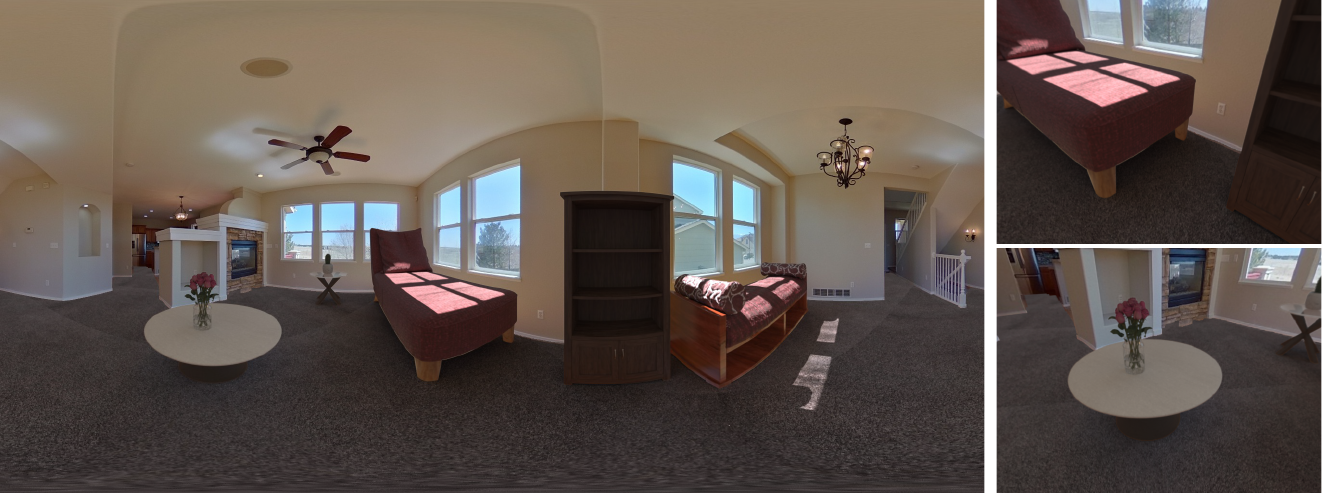}}
\subfloat[(h) Change to Carpet + Change Sun Direction + Insertion]{
\includegraphics[width=0.49\linewidth]{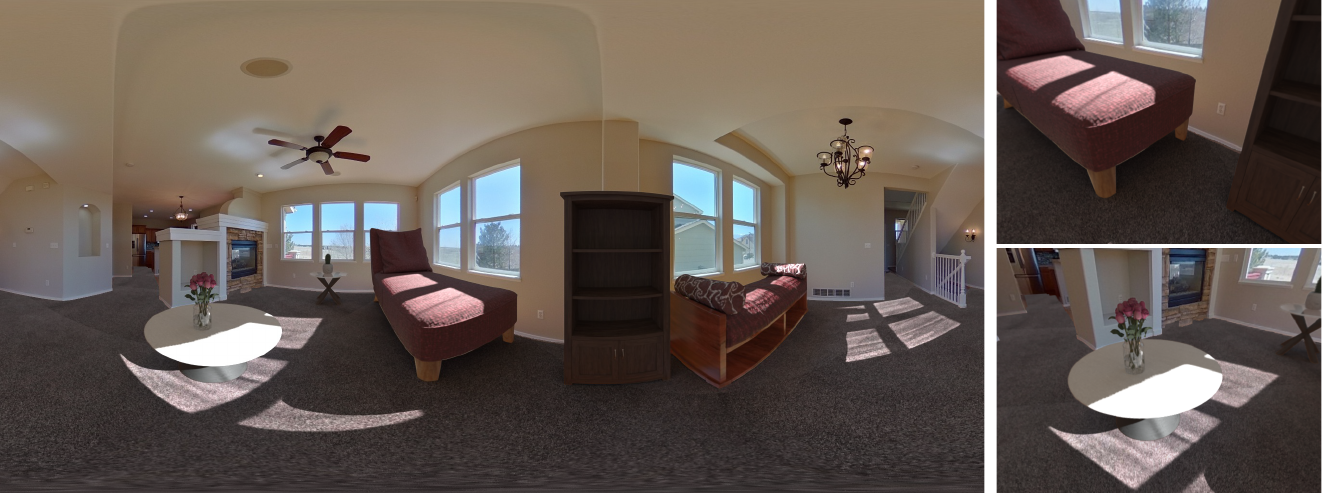}}
\caption{Appearance decomposition enables several applications such as furniture insertion, and changing flooring and sun direction, making it a useful tool for virtual staging\protect\footnotemark[8].
Top row: a home office and a bedroom staged by inserting chair, table, bed, shelves, plant and other objects. Rows 2, 3, 4: a living room is staged with different furniture, multiple new flooring (dark wood, carpet), and is shown under two different sun directions. Notice the sun light on the surfaces and the blocking of specular reflection that add realism to the scene. }
\label{fig:apps}
\end{figure*}

%% file: fig_compinsert.tex
\begin{figure}
    \centering
    \subfloat[(a) Empty]{
    \includegraphics[width=0.24\linewidth]{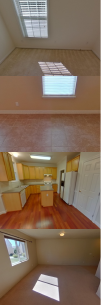}}
    \subfloat[(b) Li~\etal~\shortcite{li2020inverse}]{
    \includegraphics[width=0.24\linewidth]{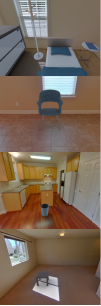}}
    \subfloat[(c) No Decomp]{
    \includegraphics[width=0.24\linewidth]{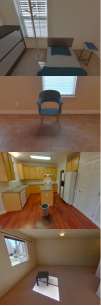}}
    \subfloat[(d) Ours]{
    \includegraphics[width=0.24\linewidth]{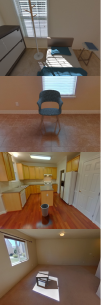}}
    \caption{We insert furniture\protect\footnotemark[9] by applying the inverse rendering approach by Li~\etal~\cite{li2020inverse} to perspective crops (a) of our panoramas (b). This method is trained on perspective views with synthetically rendered data. We also insert furniture by using the HDR map and 3D layout as illumination (c). Both approaches are not able to faithfully render the shading interactions of the objects with sunlight or specular reflections (see ours in (d)).}
    \label{fig:compinsert}
\end{figure}

%% file: sec9_conclude.tex
\section{Conclusion}

In summary, we present an appearance decomposition method for empty indoor panoramic scenes. Relying on  supervision from semantics, our method detects and removes the specular reflection on the floor and the direct sunlight on the floor and the walls.
The decomposition result can be applied to multiple virtual staging tasks, including albedo and sun direction estimation, furniture insertion, changing sun direction and floor material.

\footnotetext[9]{3D model credits: Row 1: See Fig.~\ref{fig:apps} (b), Row 2: armchair\copyright Dmitrii Ispolatov/Adobe Stock, Row 3: trash can\copyright adobestock3d/Adobe Stock, Row 4: table\copyright Dmitrii Ispolatov/Adobe Stock.}

\input{fig_elevated}

There are three main limitations. First, we focus on direct lighting effects (first light bounce, no global illumination) on the floor and the walls. However, the specular reflection and direct sunlight could also appear on objects like kitchen countertops or appliances. Fig.~\ref{fig:elevated} shows that our method partially detects the sunlight but sun direction and object lighting is erroneous since the surface height is unknown. 
This could potentially be solved by extending our semantics and geometry to include furniture or other objects as long as some coarse mask and geometry can be obtained automatically to allow for accurate ray-tracing. Second, we assume the window producing the sunlight is visible for sun direction estimation. Using multi-view information from different panoramas in a single home can help overcome this limitation. Third, our approach is tailored to panoramas. It is possible to extend our ideas to narrower perspective field of views but there are several alternate methods in that space and we lose the advantages provided by a panorama. Given the increased availability of panoramas of homes, we believe the method is timely and can be an effective tool for virtual staging.
\footnotetext[10]{3D model credit: mug\copyright Brandon Westlake/Adobe Stock.}

%% file: fig_elevated.tex
\begin{figure}
    \centering
    \subfloat[(a) RGB]{
    \includegraphics[width=0.49\linewidth]{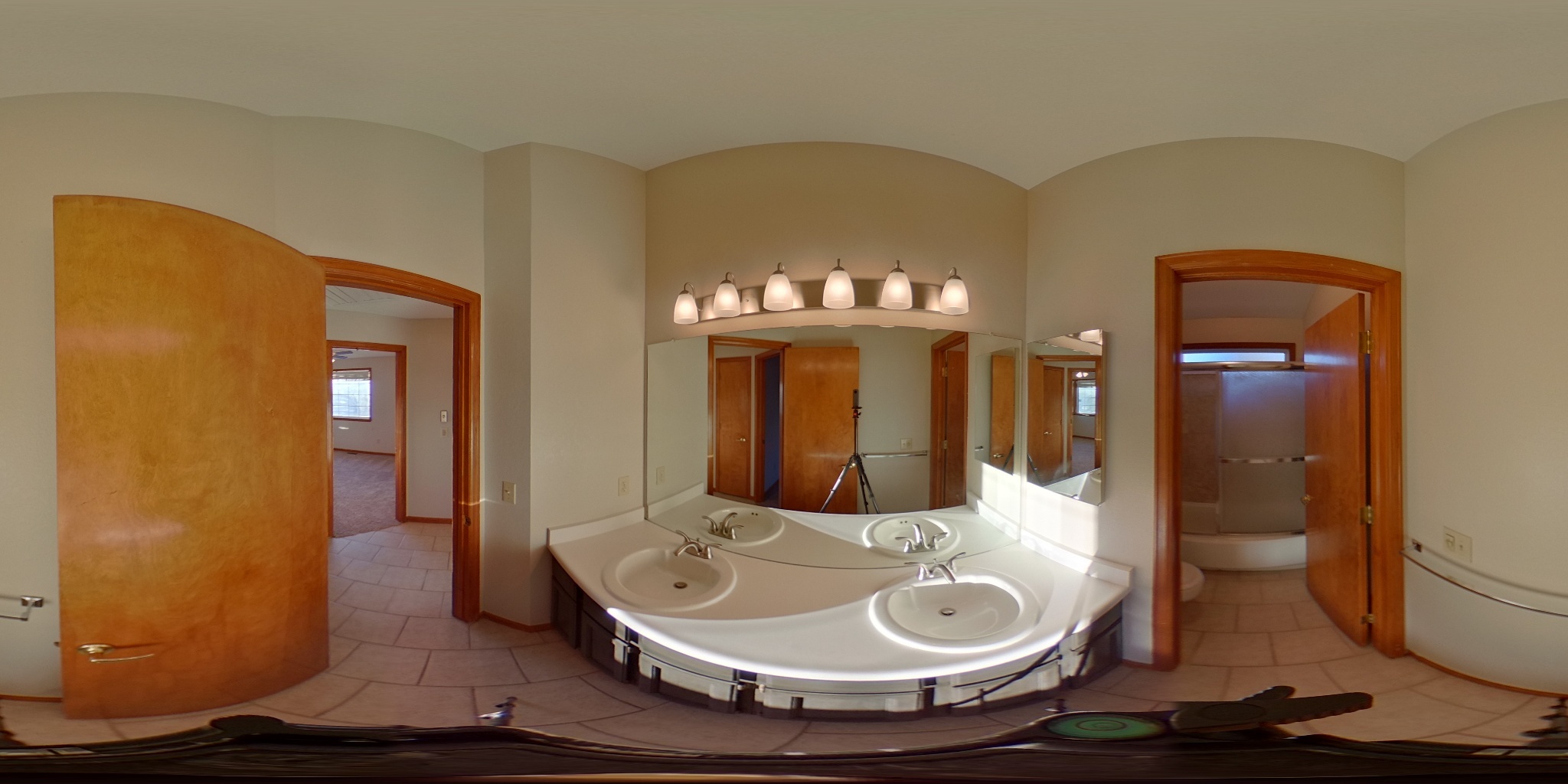}}
    \subfloat[(b) Lighting Effects (brightened)]{
    \includegraphics[width=0.49\linewidth]{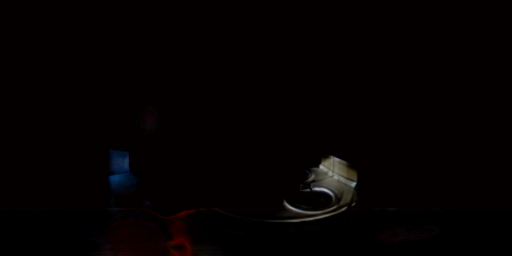}}
    
    \subfloat[(c) Object Insertion]{
    \includegraphics[width=0.49\linewidth]{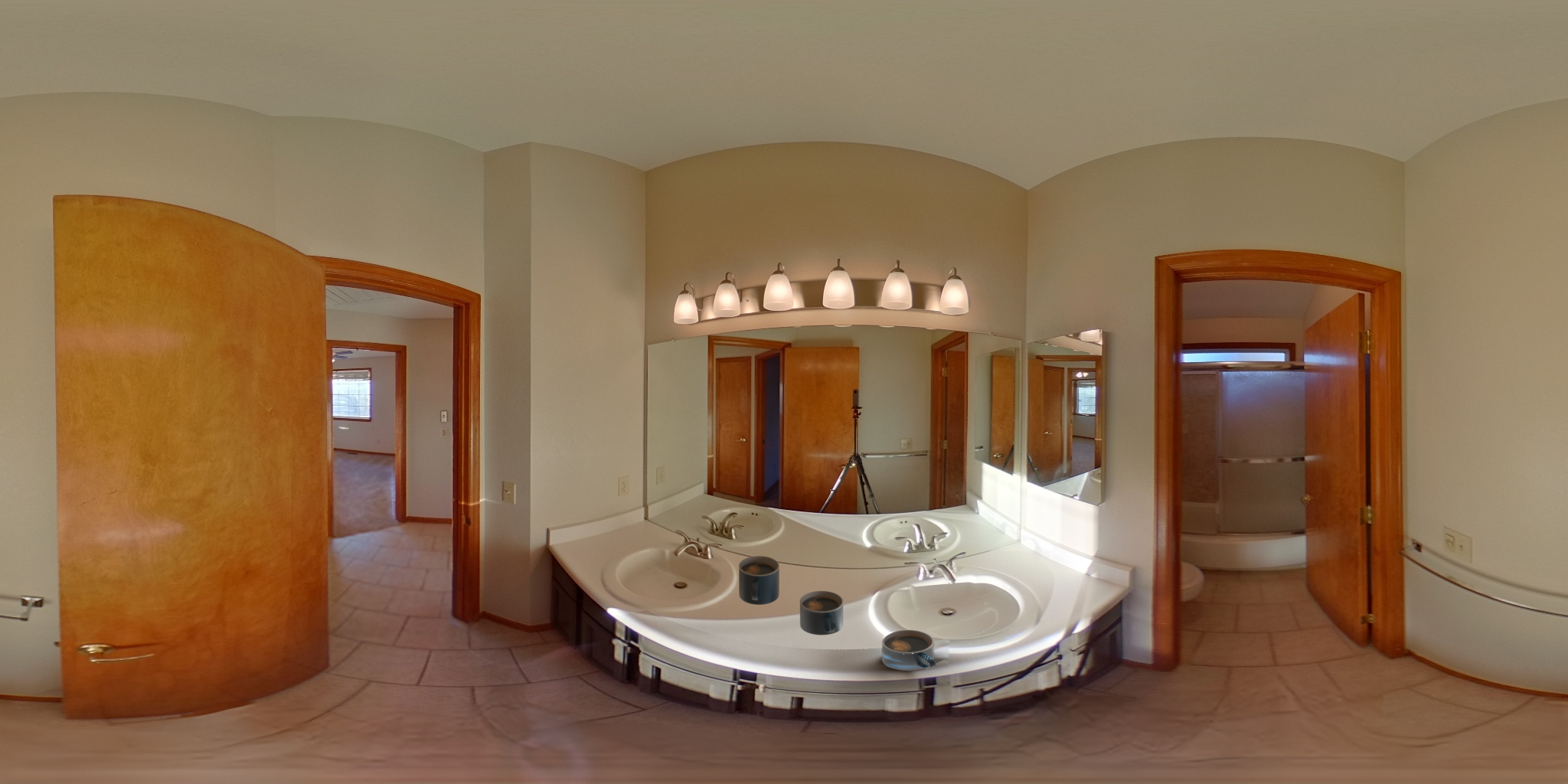}}
    \subfloat[(d) Perspective View]{
    \includegraphics[width=0.49\linewidth]{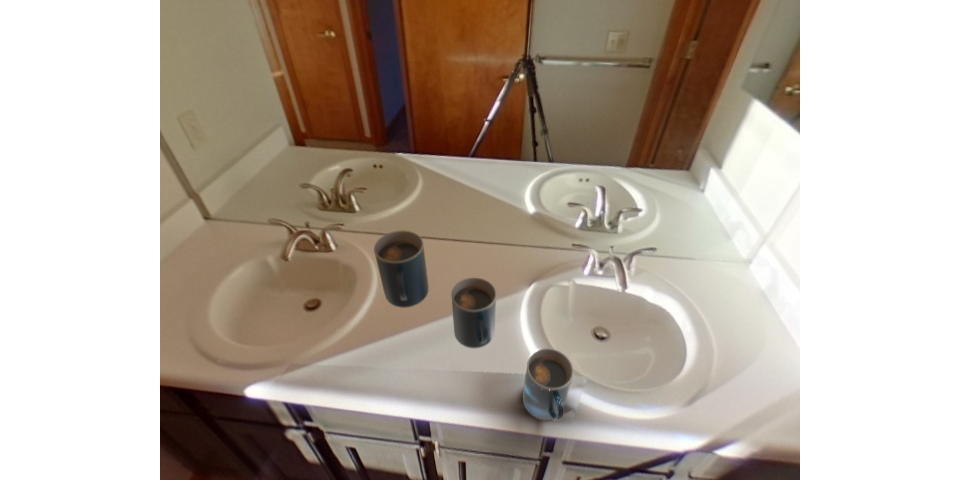}}

    \caption{Failure case: elevated surface. Our method partially detects the sunlight but the sun direction and object lighting\protect\footnotemark[10] is erroneous, because of incorrect geometry estimation (surface with unknown geometry is treated as floor or wall). Although the mug in the sunlit region is brighter, the shadows are not correctly rendered.}
    \label{fig:elevated}
\end{figure}